\documentclass [twocolumn]{aastex62}
\usepackage{hyperref} 
\usepackage{amssymb, amsmath, mathtools}
\usepackage{natbib}
\usepackage{enumerate}
\usepackage{graphicx}

\graphicspath{{./}{Figures/}}

\shorttitle{Sample article}
\shortauthors{Schwarz et al.}

\begin{document}

\title{Transmission Spectroscopy of WASP-79b from 0.6 to 5.0 $\mu$m}
\markright{Transmission Spectroscopy of WASP-79b from 0.6 to 5.0 $\mu$m , K.S. Sotzen et al}

\correspondingauthor{Kristin Showalter Sotzen}
\email{kristin.sotzen@jhuapl.edu, kshowal3@jhu.edu}

\author[0000-0001-7393-2368]{Kristin S. Sotzen}
\affil{Johns Hopkins University, 3400 N. Charles St, Baltimore, MD 21218, USA}
\affiliation{ JHU Applied Physics Laboratory, 11100 Johns Hopkins Rd, Laurel, MD 20723, USA}

\author{Kevin B. Stevenson}
\affiliation{ JHU Applied Physics Laboratory, 11100 Johns Hopkins Rd, Laurel, MD 20723 USA}
\affiliation{Space Telescope Science Institute, 3700 San Martin Dr, Baltimore, MD 21218, USA}

\author{David K. Sing}
\affil{Johns Hopkins University, 3400 N. Charles St, Baltimore, MD 21218, USA}

\author{Brian M. Kilpatrick}
\affiliation{Space Telescope Science Institute, 3700 San Martin Dr, Baltimore, MD 21218, USA}

\author{Hannah R. Wakeford}
\affiliation{Space Telescope Science Institute, 3700 San Martin Dr, Baltimore, MD 21218, USA}

\author{Joseph C. Filippazzo}
\affiliation{Space Telescope Science Institute, 3700 San Martin Dr, Baltimore, MD 21218, USA}

\author{Nikole K. Lewis}
\affiliation{Department of Astronomy and Carl Sagan Institute, Cornell University, 122 Sciences Drive, Ithaca, NY 14853, USA }

\author{Sarah M. H{\"o}rst}
\affil{Johns Hopkins University, 3400 N. Charles St, Baltimore, MD 21218, USA}
\affiliation{Space Telescope Science Institute, 3700 San Martin Dr, Baltimore, MD 21218, USA}

\author{Mercedes L{\'o}pez-Morales}
\affiliation{Center for Astrophysics ${\rm \mid}$ Harvard {\rm \&} Smithsonian, 60 Garden St, Cambridge, Cambridge, MA 02138, USA}

\author{Gregory W. Henry}
\affil{Center for Excellence in Information Systems, Tennessee State University, Nashville, TN 37209, USA}

\author{Lars A. Buchhave}
\affil{DTU Space, National Space Institute, Technical University of Denmark, Elektrovej 328, DK-2800 Kgs. Lyngby, Denmark}

\author{David Ehrenreich}
\affil{Observatoire de l’Universit{\'e} de Gen{\`e}ve, 51 chemin des Maillettes, 1290 Sauverny, Switzerland}

\author{Jonathan D. Fraine}
\affiliation{Space Science Institute, 4750 Walnut St \#205, Boulder, CO 80301, USA}

\author{Antonio Garc{\'i}a Mu{\~n}oz}
\affil{Zentrum f{\"u}r Astronomie und Astrophysik, Technische Universit{\"a}t Berlin, EW 801, Hardenbergstrasse 36, D-10623 Berlin, Germany}

\author{Rahul Jayaraman}
\affil{Brown University, Department of Physics, Box 1843, Providence, RI 02904, USA}

\author{Panayotis Lavvas}
\affil{Groupe de Spectrom{\'e}trie Moleculaire et Atmosph{\'e}rique, Universit{\'e} de Reims Champagne Ardenne, Reims, France}

\author{Alain Lecavelier des Etangs}
\affil{Institut d'astrophysique de Paris, UMR7095 CNRS, Sorbonne Universit\'e, 98bis Boulevard Arago, 75014 Paris, France}

\author{Mark S. Marley}
\affil{NASA Ames Research Center, MS 245-3, Moffett Field, CA 94035, USA}

\author{Nikolay Nikolov}
\affil{Johns Hopkins University, 3400 N. Charles St, Baltimore, MD 21218, USA}

\author{Alexander D. Rathcke}
\affil{DTU Space, National Space Institute, Technical University of Denmark, Elektrovej 328, DK-2800 Kgs. Lyngby, Denmark}

\author{Jorge Sanz-Forcada}
\affil{Centro de Astrobiolog{\'i}a (CSIC-INTA), E-28692 Villanueva de la Ca{\~n}ada, Madrid, Spain}

\received{3 September 2019}
\revised{31 October 2019}
\accepted{3 November 2019}

\begin{abstract}

As part of the PanCET program, we have conducted a spectroscopic study of WASP-79b, an inflated hot Jupiter orbiting an F-type star in Eridanus with a period of 3.66 days. Building on the original WASP and TRAPPIST photometry of \cite{Smalley2012}, we examine HST/WFC3 (1.125 - 1.650 {$\mu$}m), Magellan/LDSS-3C (0.6 - 1 {$\mu$}m) data, and \textit{Spitzer} data (3.6 and 4.5 {$\mu$}m).  Using data from all three instruments, we constrain the water abundance to be --2.20 $\leq$ log(H\textsubscript{2}O) $\leq$ --1.55. We present these results along with the results of an atmospheric retrieval analysis, which favor inclusion of FeH and H\textsuperscript{-} in the atmospheric model. We also provide an updated ephemeris based on the Smalley, HST/WFC3, LDSS-3C, Spitzer, and TESS transit times.  With the detectable water feature and its occupation of the clear/cloudy transition region of the temperature/gravity phase space, WASP-79b is a target of interest for the approved JWST Director's Discretionary Early Release Science (DD ERS) program, with ERS observations planned to be the first to execute in Cycle 1.  Transiting exoplanets have been approved for 78.1 hours of data collection, and with the delay in the JWST launch, WASP-79b is now a target for the Panchromatic Transmission program.  This program will observe WASP-79b for 42 hours in 4 different instrument modes, providing substantially more data by which to investigate this hot Jupiter.

\end{abstract}

\keywords{methods --- observational{$:$} 
atmospheres --- planets and satellies{$:$} individual --- WASP-79b}


\section{Introduction} \label{sec:intro}

Based on studies of planets and moons within the solar system and spectral analyses of exoplanets, a persistent atmosphere is generally accompanied by clouds and/or hazes. Recent studies of hot Jupiters have revealed that many of the exoplanets observed in transmission have cloudy or hazy properties, with their spectra dominated by strong optical Rayleigh and/or Mie scattering from high-altitude aerosol particles (e.g., \cite{Sing2016, Stevenson2016a, Wakeford2015, Lavvas2017}). Clouds and hazes in exoplanetary atmospheres can have a significant impact on the detectable spectra for these worlds. In the optical range, small particles produce scattering that leads to steep slopes that progressively become shallower as the particle radius increases (see e.g., \cite{Lavvas2017}). This scattering effectively dampens any features from the deeper atmosphere, including pressure-broadened alkali Na and K lines, and can mute or obscure expected water absorption features in the near-infrared (see e.g., \cite{Wakeford2015}).

The majority of current exoplanet spectra are constructed from wavelengths in the optical and near-infrared wavelengths, revealing information on the portion of transmission spectra for aerosols where only scattering features are seen. When interpreting these observations, the slope of spectra in the optical regime is proportional to the temperature of the atmosphere and can be indicative of specific species when small grain sizes are considered \citep{Wakeford2015}.  Additionally, absorption features in the near- and mid-infrared spectra can be identified as the vibrational modes of the major bond pairs in certain potential condensates, providing composition information \citep{Wakeford2015}.

The survey analysis performed by \cite{Sing2016} of ten hot Jupiters found that planets with predominantly clear atmospheres show prominent alkali and H\textsubscript{2}O absorption, with infrared radii values commensurate or higher than the optical altitudes, while heavily hazy and cloudy planets have strong optical scattering slopes, narrow alkali lines, and H\textsubscript{2}O absorption that is partially or completely obscured.

Like many transiting exoplanets found using ground-based surveys, WASP-79b is a hot Jupiter with an extended atmosphere.  Discovered in 2012 by Smalley et al using photometry from the WASP-South and TRAPPIST telescopes, it was found to have a planetary mass of 0.90 {$\pm$} 0.08 M\textsubscript{$Jup$} and a large radius estimate, ranging from 1.7 {$\pm$} 0.11 R\textsubscript{$Jup$} using a main-sequence mass-radius constraint on the Markov Chain Monte Carlo (MCMC) process, to 2.1 {$\pm$} 0.14 R\textsubscript{$Jup$} using a non-main sequence constraint \citep{Smalley2012}. While both radius estimates were large for the available hot Jupiter data in 2012, the estimate based on the non-main sequence constraint would have made WASP-79b the largest exoplanet discovered at the time \citep{Smalley2012}. With a mass estimate of approximately one M\textsubscript{$Jup$} and such a large radius estimate, WASP-79b's density is comparatively low, implying that its atmosphere is extended.  In addition, the host star WASP-79 is a bright, quiet F-type star with consistent stellar activity, with variation in the baseline stellar flux within 0.1{$\%$} (Section \ref{sec:results}).  

WASP-79b has a $T_{eq}$ $\sim$1800 K and a log $g$ between 2.67 and 2.85 \citep{Smalley2012}, placing this planet in a transition region of the temperature/gravity phase space.  On one side of this transition region, planets have been found to have muted water features due to clouds and hazes, while on the other side, planets have been found to have strong measured water features, implying clearer atmospheres \citep{Stevenson2016b}.  Being in this transition region, WASP-79b provided an opportunity to further study this relationship between temperature, gravity, and the presence of atmospheric clouds and/or hazes.  These studies are important for predictions of atmospheric feature obscuration, which inform target selection and observations for telescopes like the Hubble Space Telescope (HST).




Additionally, with its broad observing windows \citep{Bean2018}, WASP-79b presented an excellent candidate for a transmission spectroscopy study as well as a potential Early Release Science (ERS) candidate for the James Webb Space Telescope (JWST).  It was therefore scheduled for follow-up observations using HST, the Magellan Large Dispersion Survey Spectrograph 3 (LDSS3), and the \textit{Spitzer} Space Telescope to determine its value as a candidate for JWST observation, with broad wavelength coverage to evaluate its value as an ERS candidate.

In Sections \ref{sec:tess}, \ref{sec:hstObs}, \ref{sec:ldss3Obs}, and \ref{sec:spitzer}, we describe observations, analysis methods, and results from TESS, HST, LDSS3, and Spitzer respectively. In Section \ref{sec:discuss}, we discuss the atmospheric retrieval analysis and expectations for JWST observations, and in Section \ref{sec:conc}, we present our conclusions.

\section{Observations}

\subsection{TESS Data} \label{sec:tess} 
The Transiting Exoplanet Survey Satellite (TESS) observed 12 transits of WASP-79b in January and February of 2019.  TESS provides data in the 0.6 - 1.0 $\mu$m band, and the TESS light curve contains data covering 12 transits in Sectors 4 and 5. 
We fit the TESS WASP-79b 2-minute cadence transits using the Presearch Data Conditioning (PDC) light curve, which has been corrected for effects such as non-astrophysical variability and crowding \citep{2016SPIE.9913E..3EJ}.
From the timeseries, we removed all of the points which were flagged with anomalies.  The Barycentric TESS Julian Dates (BTJD) were converted to BJD$_{\rm TDB}$ by adding 2,457,000 days. 
For each transit in the light curve, we extracted a 0.5 day window centered around the transits and fit each transit event individually.
We fit the data using the 4-parameter non-linear limb-darkened transit model of \cite{Mandel2002} and included a linear baseline time trend. We calculated the limb-darkening coefficients as in \cite{Sing2010} using a Kurucz stellar model finding coefficients of $c_1=0.5012$, $c_2=0.2630$, $c_3=-0.1034$, and $c_4=-0.0301$.
For each of the 12 transits, we fit for six free parameters consisting of the central transit time, planet-to-star radius ratio, linear baseline, $cosi$, and $a/R*$.  
The high-quality of the TESS transit light curves places tight constraints on the system parameters, and we find
a weighted-average inclination of $i$=85.929$\pm$0.174 degrees and $a/R*$=7.292$\pm$0.080.  These planetary parameters were used as fixed values in the HST, LDSS3, and Spitzer analyses.
Fixing the system parameters with these values for use in the transmission spectra, we find a weighted-average value of $R_{pl}(\rm{TESS})/R_{star}=0.10675\pm0.00014$, which is in good agreement with the HST, Spitzer, and LDSS values.


\begin{figure}[t]
\includegraphics[width=1.0\linewidth]{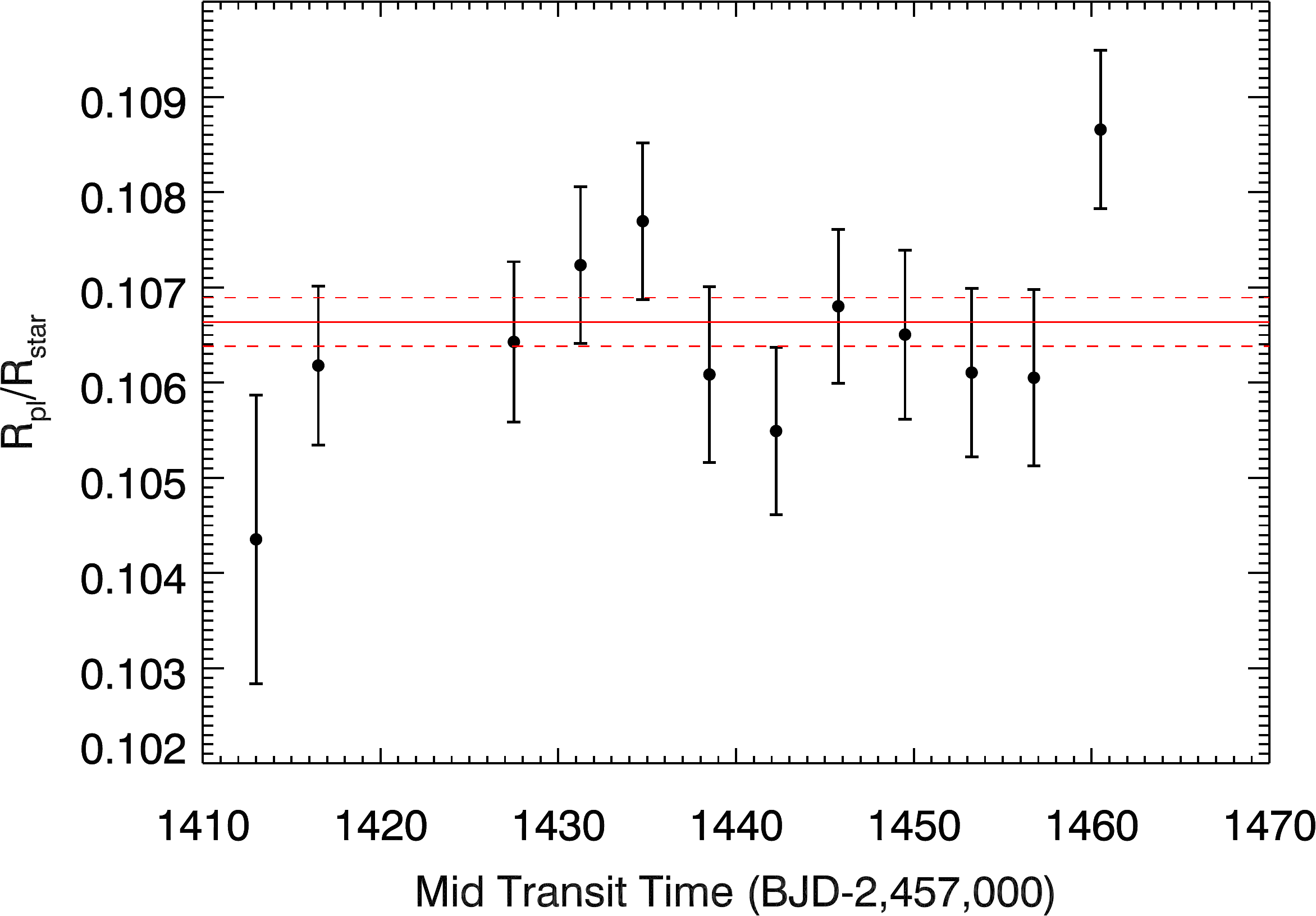}
\caption{\label{fig:tess_rprs}\small
Transit depth estimates for the 12 transits of WASP-79b available from TESS.  Estimates are shown with 1{$\sigma$} uncertainties.  The red lines indicate the weighted mean of the transit depths with 1{$\sigma$} uncertainties.}
\end{figure}

\subsection{HST/WFC3 Observations and Data Analysis} \label{sec:hstObs}

\subsubsection{Observations} \label{subsec:wfc3Obs}
We analyzed WASP-79b WFC3 data from the Panchromatic Exoplanet Treasury (PanCET) program (HST GO-14767, P.I.s Sing \& L{\'o}pez-Morales).  During its primary transit in March of 2017, HST observed WASP-79b in spatial scan mode, which slews the telescope during the exposure and moves the spectrum perpendicularly to the dispersion direction on the detector \citep{Kreidberg2014}.  This mode allows for longer integration times by distributing the incoming energy over multiple pixels. The Wide Field Camera 3 (WFC3) instrument utilized its G141 GRISM to acquire spectra from 1.1 to 1.7 {$\mu$}m over 5 HST orbits, during which we collected 65 science frames using 138-second integrations.  We provide an overview of the data analysis process below, and a detailed description of the process can be found in \cite{Stevenson2014a}.


\subsubsection{Reduction, Extraction, and Calibration of Spectra} \label{subsec:wfc3Reduct}
The Transit Reduction, Extraction, and Calibration Software (T-RECS) pipeline produces multi-wavelength, systematics-corrected light curves from which we derive wavelength-dependent transit depths with uncertainties \citep{Stevenson2014a}.  The bias correction is performed using a series of bias frames stacked to form a single master bias frame that is applied uniformly to all of the science frames.  We extract a pixel window centered on the spectrum that includes pixels along the spatial direction that are used in the optimal spectral extraction as well as in the background subtraction \citep{Stevenson2014a}.  We modeled the spectroscopic flat field using the coefficients provided in the updated flat field file \textit{$sedFFcube-both.fits$}.  

Because the background for HST is consistent over time, areas outside of the spectrum can be used to interpolate the background values for the region within the spectrum by computing the median of each column.  We perform 5{$\sigma$} outlier detection by stacking the images in time and evaluating each pixel along the time axis for outliers.  To account for imprecision in the instrument pointing during data collection, each spectrum is cross-correlated with the first spectrum to measure and correct for the pointing drift over time \citep{Stevenson2014a}.

\subsubsection{White Light Curve Fits} \label{subsec:hstWlcFits}
The raw transit light curves for WASP-79b exhibit ramp-like systematics comparable to those seen in previous WFC3 data.  Following standard procedure for HST transit light curves, we did not include data from the first orbit in our analysis \citep{Kreidberg2014}. We corrected for systematics in the remaining orbits by modeling the systematics as a function of time, which includes an exponential ramp term fitted to each orbit, a linear trend term, and a quadratic term for limb-darkening.

\begin{figure}[t]
\includegraphics[width=1.0\linewidth]{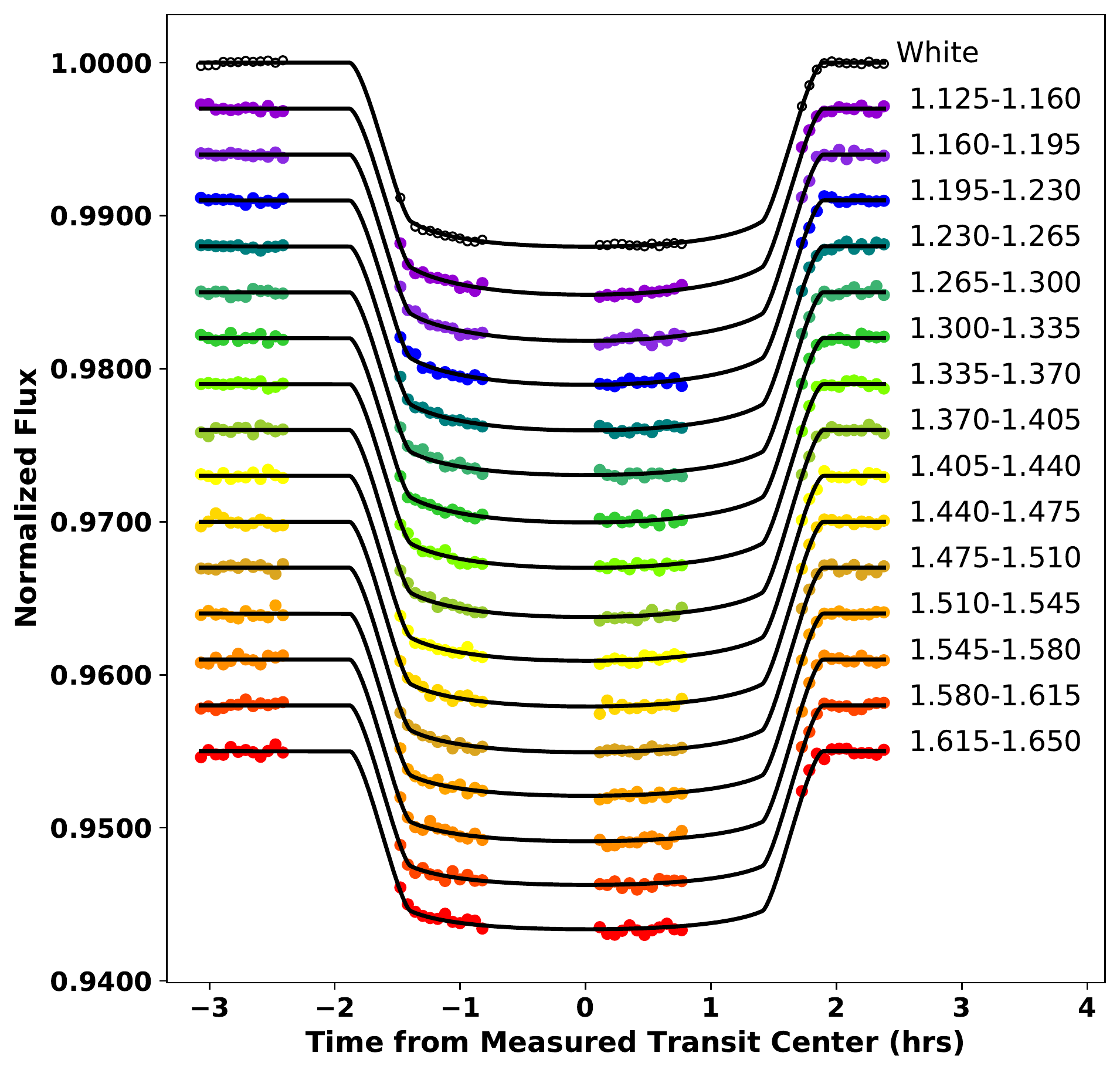}
\caption{\label{fig:hst_stacked}\small
WASP-79b white and spectroscopic light curves extracted from the HST/WFC3 data using the process described in \cite{Stevenson2014a}.  The results are binned, normalized to the system flux, and vertically shifted for ease of comparison. The error bars represent 1{$\sigma$} uncertainties.  The black lines show the best-fit models, and the wavelength range for each of the 15 channels is specified in {$\mu$}m \citep{Stevenson2014a}.}
\end{figure}

We modeled the band-integrated light curve in order to identify and remove systematics, most of which are wavelength-independent with WFC3, and to establish the absolute transit depth when comparing transmission spectra from different instruments using non-overlapping wavelengths \citep{Stevenson2014a}. We created this ``white'' light curve (WLC) by summing the flux values over the entire wavelength range. We used the Bayesian Information Criterion (BIC) to select the best systematics model component, and our final analytic model for the HST/WFC3 data takes the form:

\begin{equation}
F(t) = F\textsubscript{$s$}T(t)L(t)H(t)
\end{equation}

where \textit{$F(t)$} is the measured flux at time \textit{$t$}; \textit{$F\textsubscript{$s$}$} is the out-of-transit system flux; \textit{$T(t)$} is the primary-transit model component with unity out-of-transit flux \citep{Mandel2002}; \textit{$L(t) = a(t-t\textsubscript{$0$}) + 1$} is the time-dependent linear model component with a fixed offset, \textit{$t\textsubscript{$0$}$}, and free parameter, \textit{a}; and \textit{$H(t) = 1 - exp(-a\times{}P + b) + c\times{}P$} fits the HST ``hook'' using a rising exponential with free parameters \textit{a}, \textit{b}, where \textit{c}, and P represents the number of \textit{HST} orbits since the beginning of the transit.  The white light curve extraction for the HST/WFC3 data resulted in a transit depth of 1.1282{$\%$} {$\pm$} 0.0032{$\%$} (see Figure \ref{fig:hst_stacked}).

\begin{figure}[t]
\includegraphics[width=1.0\linewidth]{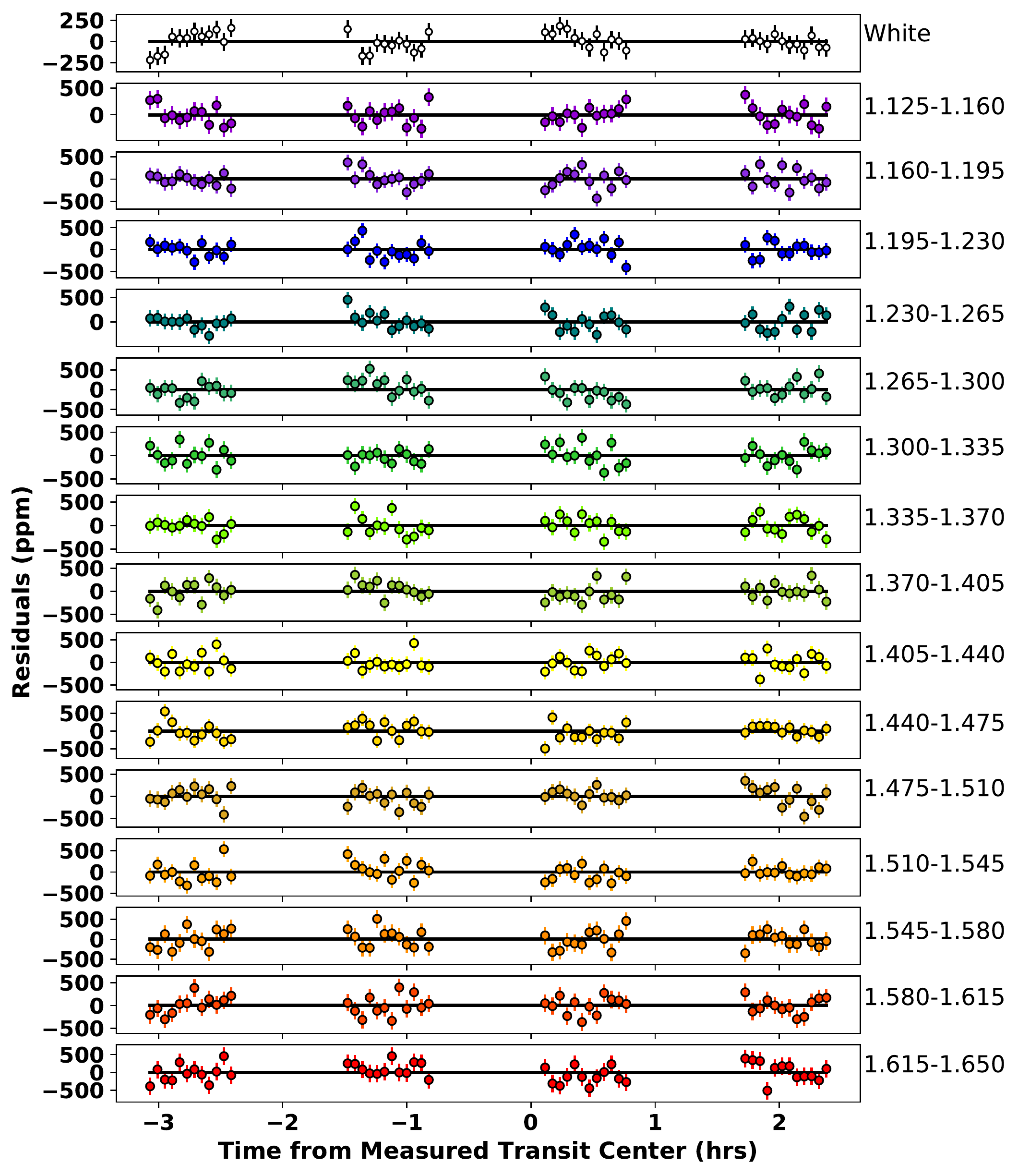}
\caption{\label{fig:hst_stacked_res}\small
White and spectroscopic residuals for light curves extracted from the HST/WFC3 data.  Values represent 1{$\sigma$} residuals.}
\end{figure}

\subsubsection{Light Curve Fits} \label{subsec:hstLcFits}
We use the \textit{Divide-White} method described by \cite{Stevenson2014a} to model the wavelength-dependent (i.e., spectroscopic) light curves, without making any prior assumptions about the form of the systematics, by utilizing information within the wavelength-independent (white) light curves.  This can be done for an arbitrary number of wavelength bins, though ten to fifteen bins provides sufficient resolution to reveal features of interest while maintaining sufficient signal to noise in each bin.

To construct a spectrum, we are interested only in the relative transit depths of the different wavelength bins.  We can therefore estimate uncertainties with our Differential Evolution Markov Chain Monte Carlo (DE-MCMC) algorithm, assuming fixed parameters for \textit{a/R*} and cos\textit{i} \citep{Stevenson2014a}.   For the HST, LDSS-3C (Section \ref{sec:ldss3Obs}), and Spitzer (Section \ref{sec:spitzer}), we assumed a fixed \textit{a/R*} of 7.2900 and a \textit{cosi} of 0.070993, based on an analysis of the TESS data for WASP-79b (Section \ref{sec:tess}).  The transit midpoint was carried as a free parameter and estimated in the WLC analyses and then fixed for the spectroscopic analyses, as it is wavelength-independent.  Figure \ref{fig:hst_stacked} shows results for the HST white light curve extraction as well as results for the 15 wavelength bins from the spectroscopic light curve extraction.


The results of the HST/WFC3 analysis, which indicate the presence of water in WASP-79b's atmosphere, are discussed in later sections.

\subsection{LDSS-3C Observations and Data Analysis} \label{sec:ldss3Obs}
In order to obtain a more complete picture of WASP-79b's atmospheric structure and to assess the slope (if any) of the spectrum, we extended our analysis for this planet to the visible and near-infrared using data from the Low Dispersion Survey Spectrograph (LDSS) optical imaging spectrograph on the 6.5 m Magellan II (Clay) Telescope at Las Campanas Observatory (LCO) in Chile.  We used the LDSS-3C VPH-red grism (bandpass 0.6 - 1.0 {$\mu$}m), which extended our spectral analysis of WASP-79b into the visible wavelengths where we expected to encounter the effects of Rayleigh scattering due to aerosols.

Our reduction, calibration, white light curve fitting, and spectroscopic light curve fitting processes use the T-RECS analysis pipeline and match the processes described in detail in \cite{Stevenson2016a}. We will therefore only discuss details pertaining to this specific observation set. 

\subsubsection{Observations, Reduction, and Calibration} \label{subsec:ldss3ObsRed}
We observed the primary transit of WASP-79b on the night of 2016 Dec 20 for nearly 8 hours (00:31 - 08:14 UT, airmass = (1.1 -- 1.0 -- 1.8)  (\ref{fig:Rotation_shift})), collecting 1230 science frames using 7-second integrations. We utilized LDSS-3C's turbo read mode with low gain and applied 2x2 pixel binning to minimize readout times, overall achieving a duty cycle of 31{$\%$}.  The most recent upgrade of the instrument to LDSS-3C constituted an upgrade to a deep-well detector that eliminated the fringing issues seen previously \citep{Stevenson2016c}.   

Our science masks utilized three, 12{$"$}-wide slits for observations of our target star (WASP-79, V = 10.1) and the two comparison stars (V = 10.8, 12.7). The brighter comparison star is a G dwarf star with a T\_{eff} of 5834 K. The spectra from the dimmer comparison star were too noisy to provide reliable atmospheric corrections, so we relied strictly on the brighter reference star. Unfortunately, the brighter reference star was sufficiently displaced from the target star on the detector (146.7 arcsec) that the resulting atmospheric corrections are not necessarily consistent.  This results in relatively large error bars on the transit depth estimates for the LDSS-3C data.

\subsubsection{White Light Curve Fits} \label{subsec:ldss3WlcFits}
\begin{figure}[t]
\includegraphics[width=1.0\linewidth, height=330pt]{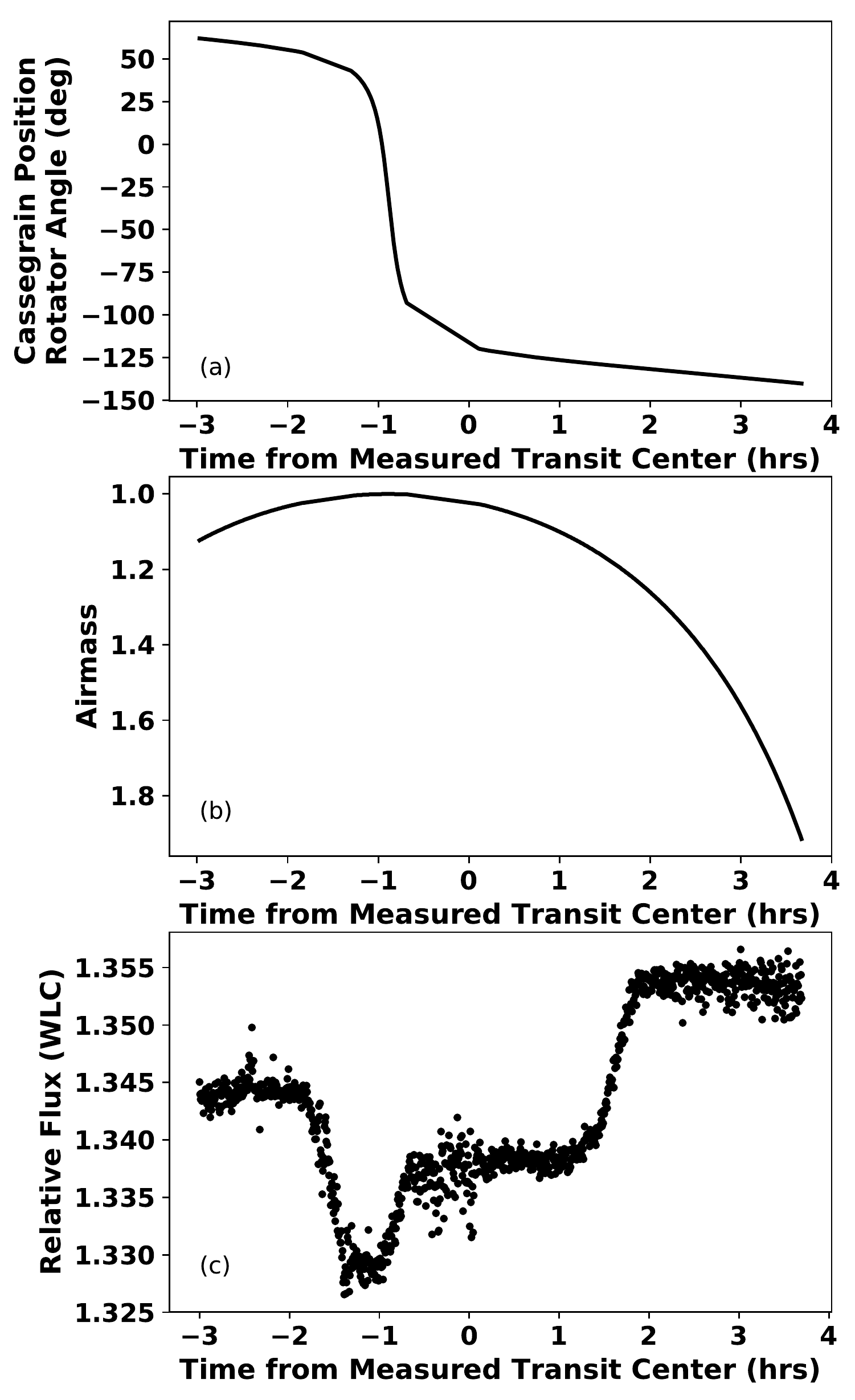}
\caption{\label{fig:Rotation_shift}\small
(a) The Cassegrain Position Rotator Angle as a function of time for the white light curve (WLC) transit extraction.  Note that the telescope passed through zenith, as indicated by both the telescope position and the airmass (b).  This resulted in a shift of the illumination on the detector and an associated shift in the relative flux between the target and reference stars, as shown in (c).}
\end{figure}

As described in \cite{Stevenson2016a}, we correct for the observed flux variations caused by fluctuations in Earth's atmosphere by dividing the WASP-79b light curve by the comparison star. We start by fitting the white light curve (0.625 - 1.025 {$\mu$}m) to maximize the signal-to-noise ratio (SNR), using both transit and systematics model components. The first utilizes a \cite{Mandel2002} transit model with selected free parameters and fixed quadratic limb-darkening parameters derived from stellar Kurucz models \citep{Kurucz2004} assuming a stellar temperature of 6500 K and log \textit{g} of 4.2. We found early in the analysis that there was a shift of the illuminated pixels on the detector in the middle of the transit that was caused by the telescope rotating as it passed through zenith (Figure \ref{fig:Rotation_shift}). For the systematics component, we tested various combinations of linear and quadratic models in combination with rotation and intra-pixel functions to account for the aforementioned rotation and pixel shift to determine which combination of models provided the best fit, based on the BIC and {$\chi$}$^{2}$ values.  Our final analytical model takes the form:

\begin{equation}
F(t) = F\textsubscript{$s$}T(t)R(t)Q(t)I(t)
\end{equation}

where \textit{$F(t)$} is the measured flux at time \textit{$t$}; \textit{$F\textsubscript{$s$}$} is the out-of-transit system flux; \textit{$T(t)$} is the primary-transit model component with unity out-of-transit flux; $R(t) = 1+ aA + b$cos$({\pi}/180\times{}({\theta}(t) + {\theta}\textsubscript{0}))$ is the time-dependent instrument rotation model component with free parameters $a$, $b$, and {$\theta$}\textsubscript{$0$}, where A = airmass; $Q(t)$ uses a quadratic polynomial to fit a pixel response ramp in the data; and $I(y)$ fits the pixel shift using a linear function in the dispersion direction.  The white light curve for the Dec 2016 LDSS-3C data resulted in a transit depth of 1.1626{$\%$} {$\pm$} 0.0152{$\%$}.


\subsubsection{Light Curve Fits} \label{subsec:ldss3LcFits}
\begin{figure}[t]
\includegraphics[width=1.0\linewidth]{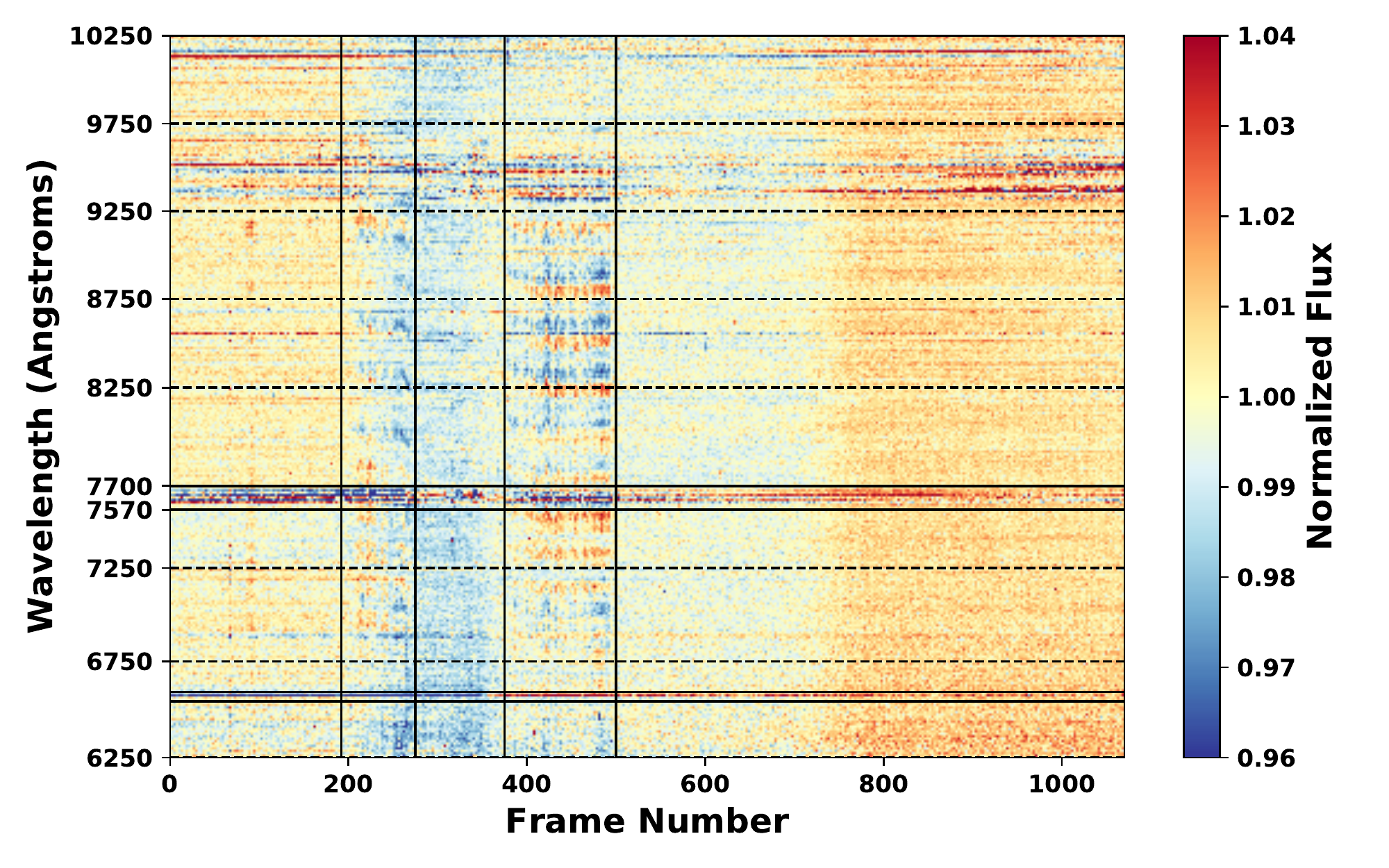}
\caption{\label{fig:ldss3_2dlc}\small
Two-dimensional light curve for the Dec 2016 LDSS-3C WASP-79b observations showing the flux of the target star normalized against the flux of the reference star.  Normalized flux is shown per wavelength as a function of frame number.  The 6540-6590, 7570-7700, and 9250-9750 channels show increased absorption, particularly early on in the observation.  The 6540-6590 and 7570-7700 channels were masked to remove their influence from the light curve extraction. The solid vertical lines indicate the times for which data was removed based on visual inspection as described in Step 1 of the iterative outlier rejection process.}
\end{figure}

As with the HST/WFC3 data, we apply the $Divide-White$ technique \citep{Stevenson2014a} to remove the wavelength-independent systematics.  To account for the wavelength-dependent systematics, each spectroscopic channel requires a rotation correction with airmass, a quadratic function in time, and an intra-pixel response shift correction.  Due to unfavorable weather effects during the night of the LDSS-3C observation, the displaced reference star, and the telescope rotation, we found the data to be very noisy with significant numbers of outliers in most channels.  To remove these outliers, we performed the following iterative outlier rejection process:

\begin{enumerate}
	\item We ran the simulation with no masking or outlier rejection so that we could visually determine whether there were any sections of the data that should be removed entirely.  Based on the results of this run and the weather information for the observation timeframe, we removed times 02:16:14 UT - 02:47:25 UT and times 03:24:58 UT - 04:11:56 UT for all channels.  Additionally, based on the normalized flux values (see Figure \ref{fig:ldss3_2dlc}), the 6540-6590 and 7570-7700 channels were masked to remove them from the light curve analysis, as they showed atmospheric absorption that could not be accounted for using the reference star, which was artificially increasing the transit depths in those channels.  There were significant changes in the local humidity over the course of the night, particularly between {$\sim$}05:00 UTC and {$\sim$}08:00 UTC that may have contributed to the noise in the data.
	\item We then ran two consecutive boxcar median masks with 3{$\sigma$} rejection on the photon flux data.
	\item We re-ran the simulation on the results from step 2, and ran three consecutive 3{$\sigma$} outlier rejection masks on the residuals for the resulting transit models.
	\item Finally, we re-ran the simulation on the results from step 3, masking the outliers identified in steps 2 and 3.
\end{enumerate}

\begin{figure*}[!htbp]
\includegraphics[height=210pt]{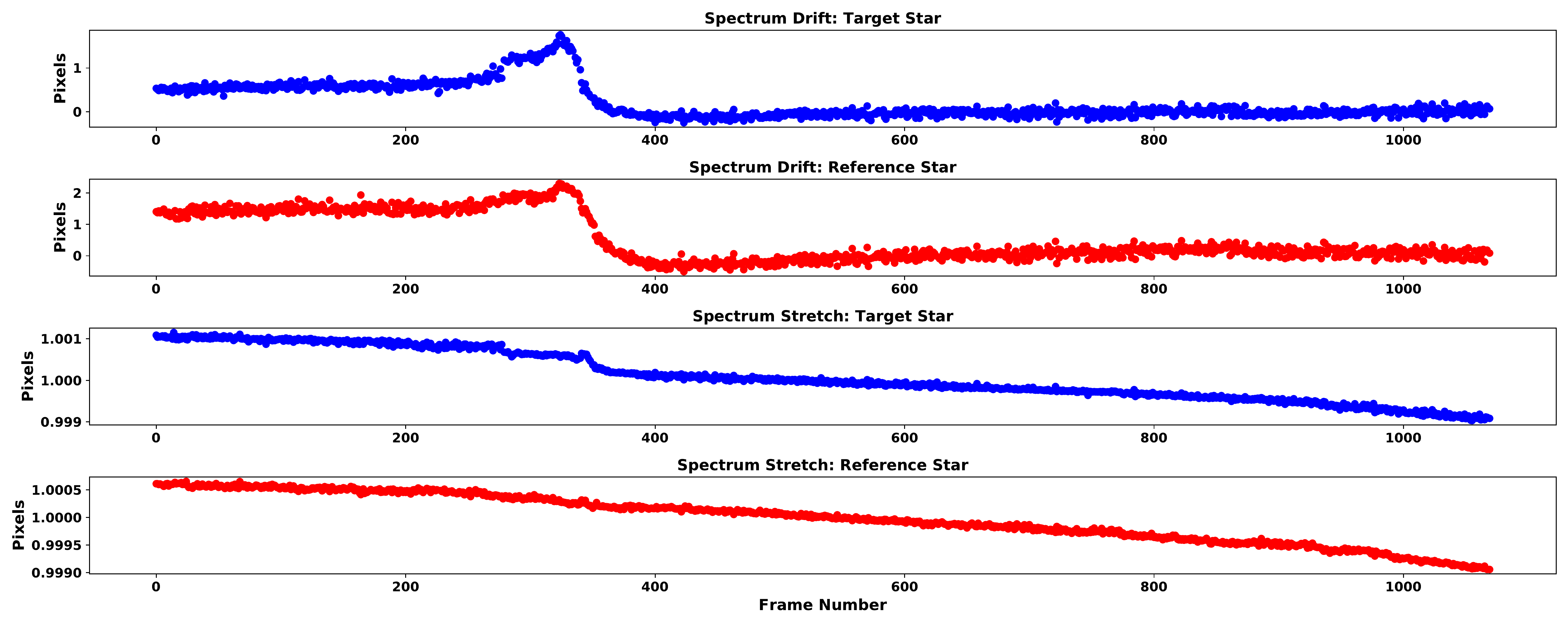}
\caption{\label{fig:ldss3_spec_drift_stretch}\small
Spectral drift in the dispersion direction and spectral stretch over the course of the observation for the target and reference stars.  The drift was in excess of 1 pixel for both the target and reference stars, while the stretch was 4 pixels for the target star over the course of the observation.}
\end{figure*}

In addition to the expected drift in the dispersion direction of the LDSS-3C spectrum over the course of the observation, \cite{DiamondLowe2018} found a stretching of the spectrum equal to approximately 4 pixels for the target star and 2 pixels for the comparison star.  To account for this effect, we calculated the stretch and the drift by optimizing a cubic spline fit of the target spectrum normalized to the reference spectrum.  Figure \ref{fig:ldss3_spec_drift_stretch} shows the calculated spectrum drift and stretch over time for both the target and reference stars, and it can be seen that the spectral drift was in excess of 1 pixel for both the target and reference stars.

\begin{figure}[!h]
\includegraphics[width=1.0\linewidth]{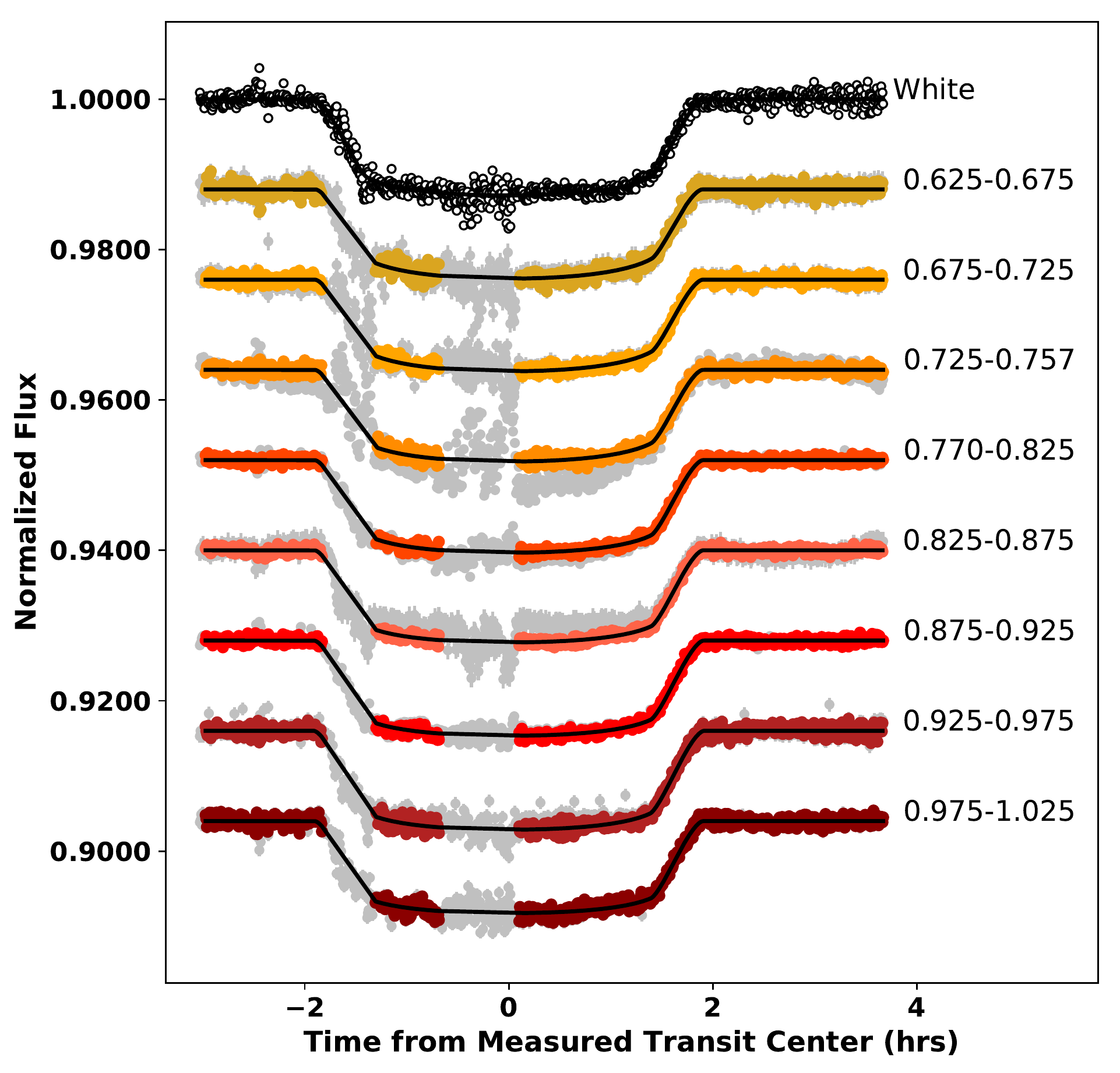}
\caption{\label{fig:ldss3_stacked}\small
WASP-79b white and spectroscopic light curves extracted from December 2016 LDSS-3C data using the fitting process described in \cite{Stevenson2016a}.  As with the WFC3 data, the results are binned and normalized to the system flux, and the error bars represent 1{$\sigma$} uncertainties.  The black lines show the best-fit models, and the wavelength range for each of the 8 channels is specified in {$\mu$}m \citep{Stevenson2016a}.  The grey points represent the original data, and the colored points represent the data that were retained from the noise and outlier masking process.}
\end{figure}

The results of the spectroscopic light curve extraction for the LDSS-3C data are shown in Figure \ref{fig:ldss3_stacked}.  Due to the large amount of noise in the data, we restricted the spectroscopic LDSS-3C analysis to 8 channels to increase the SNR.

\begin{figure}[!h]
\includegraphics[width=1.0\linewidth]{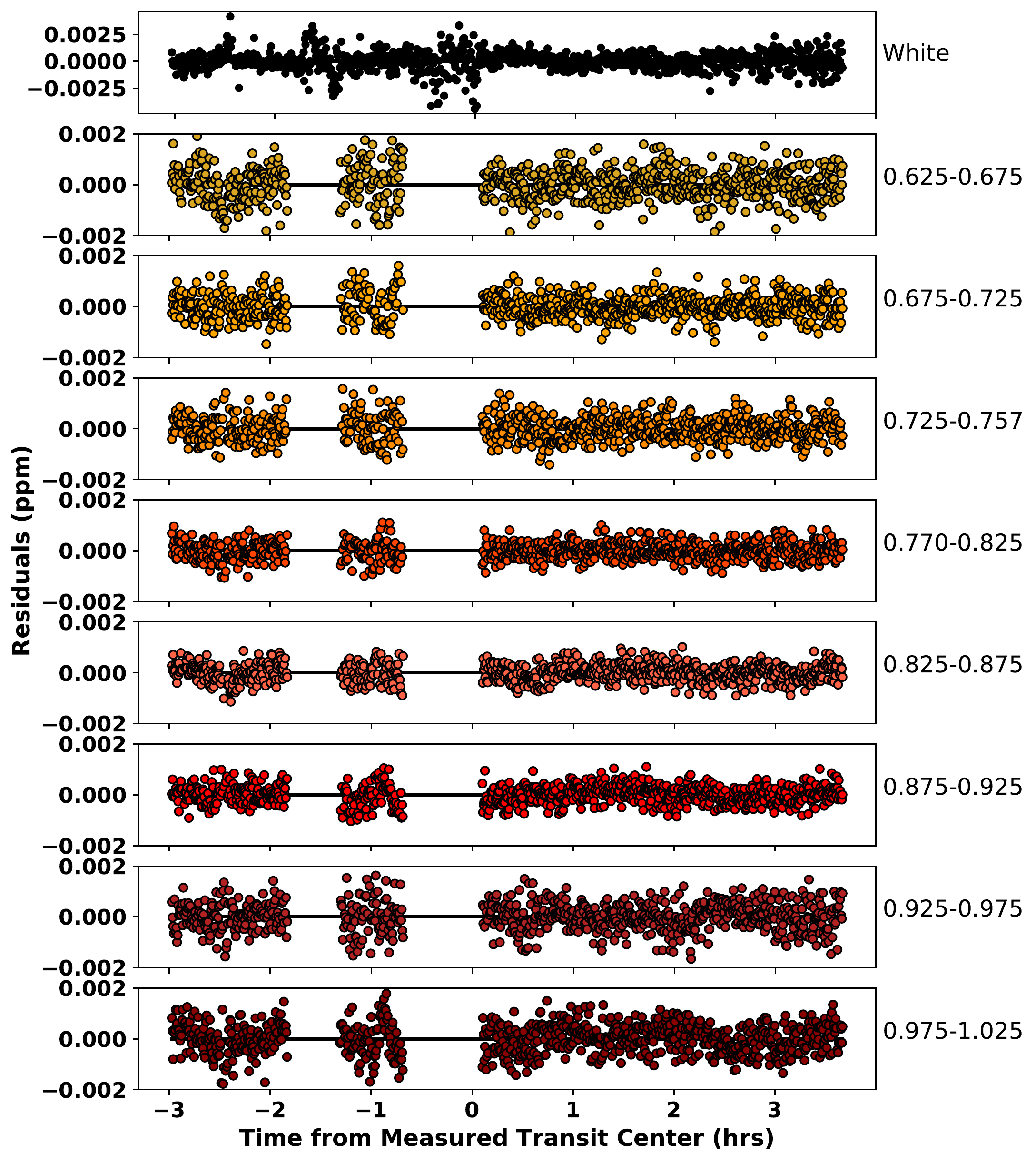}
\caption{\label{fig:ldss3_stacked_res}\small
White and spectroscopic residuals for the light curves extracted from the LDSS-3C data.  Values represent 1{$\sigma$} residuals.  The gaps in the spectroscopic plots indicate times for which noisy observation data were masked.}
\end{figure}

\subsubsection{Results} \label{sec:results}

Because the opacity of the exoplanet atmosphere varies with wavelength, the apparent size of the planet, and therefore the depth of the transit, also varies with wavelength. Having performed the spectroscopic light curve extraction and the systematics normalization via the \textit{Divide-White} method, we can construct a spectrum from the relative transit depths of the selected wavelength bins. 

\begin{figure}[t]
\includegraphics[width=1.0\linewidth]{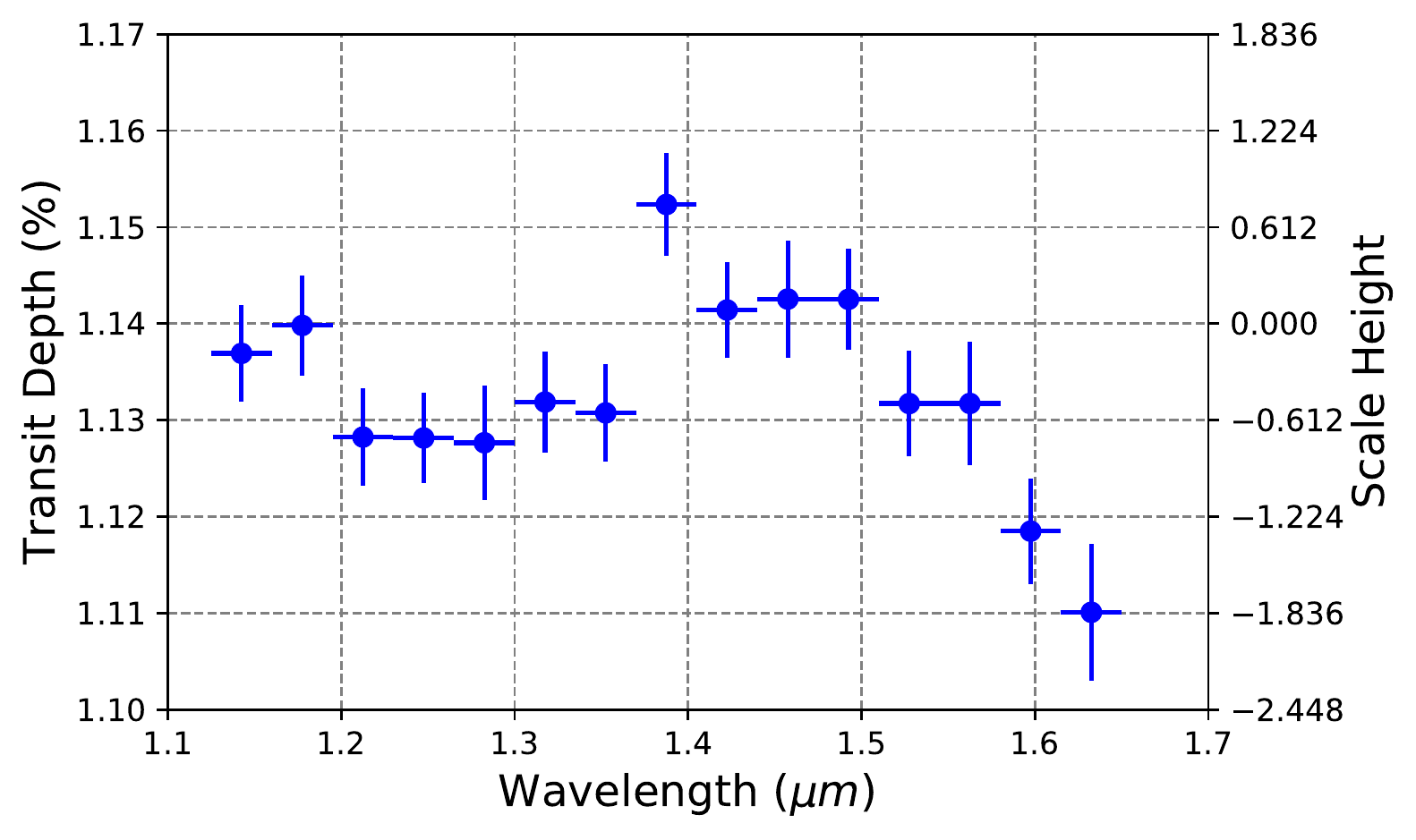}
\caption{\label{fig:hst_spectrum}\small
Spectrum constructed from transit depths of 15 wavelength bins of HST/WFC3 data. Inversion of the transit depth provides a representation of the relative absorption at different wavelengths. The increased absorption at 1.4 {$\mu$}m indicates a water absorption feature.  The horizontal error bars indicate the wavelength bins used for the light curve analysis.}
\end{figure}

Figure \ref{fig:hst_spectrum} shows the relative transit depths of the WASP-79b HST data for 15 wavelength bins for the light curve extraction using the \textit{Divide-White} normalization method. In this figure, the positive y-axis represents increasing transit depth, i.e., more absorption by the WASP-79b atmosphere. The resulting spectrum displays a noticeable peak centered at 1.4 {$\mu$}m, which represents a water feature.  This feature is consistent with water features found in the spectra of other hot Jupiters \citep{Sing2016}, and an atmospheric retrieval corroborates this feature.

\begin{figure}[t]
\includegraphics[width=1.0\linewidth]{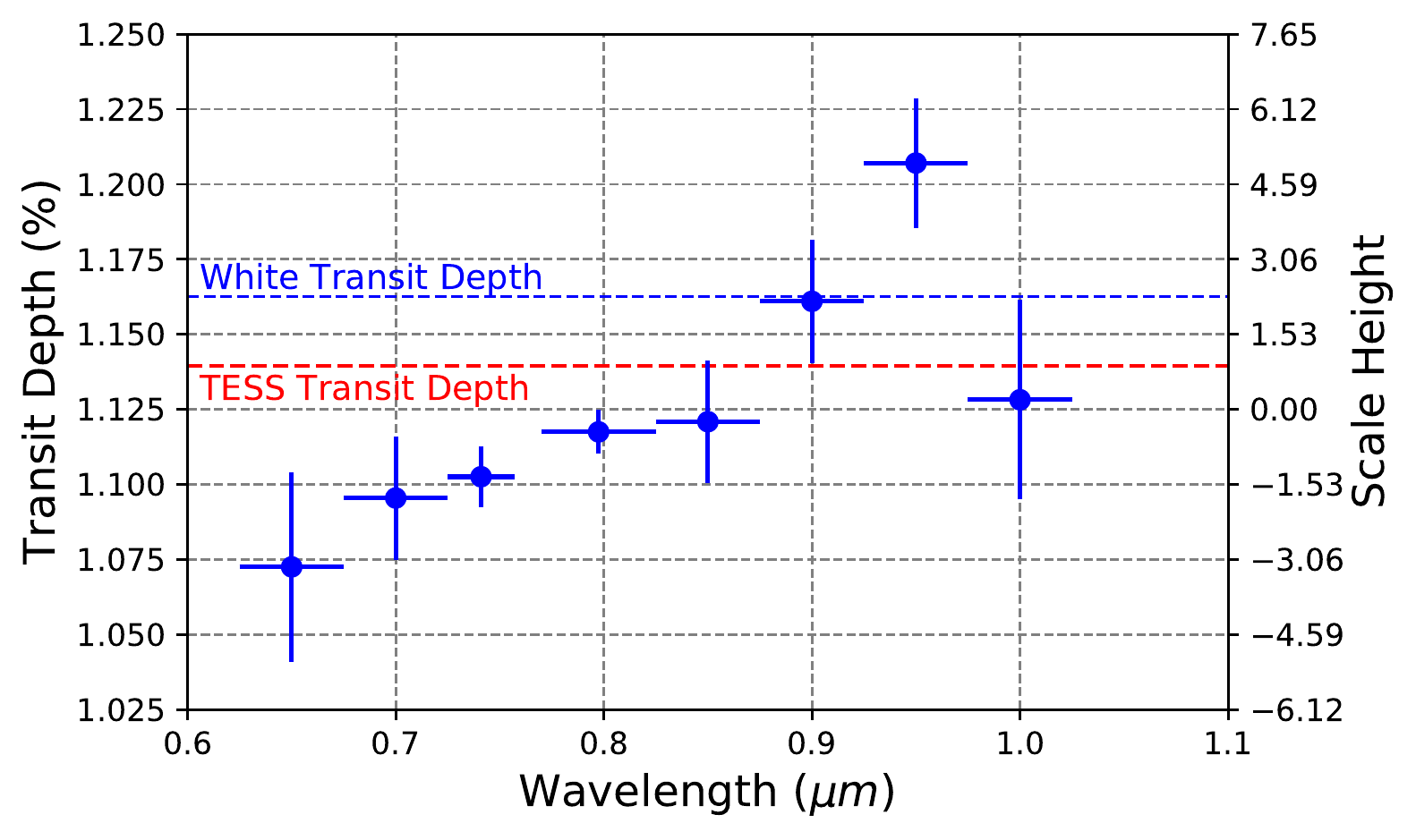}
\caption{\label{fig:ldss3_spectrum}\small
Spectrum constructed from transit depths of 8 wavelength bins of LDSS-3C data. The large spread in transit depth estimates - particularly noticeable at 0.9 and 0.95 {$\mu$}m - is likely due to interference from Earth's atmosphere that could not be fully accounted for due to the distance of the reference star from the target star.  The 0.65 {$\mu$}m point may be low due to detector cutoff at the blue edge.  The transit depth estimates for the white light curve analysis described in \ref{subsec:ldss3WlcFits} and for the TESS analysis (Section \ref{sec:tess}) are provided for comparison.}
\end{figure}

Figure \ref{fig:ldss3_spectrum} shows the relative transit depths of the WASP-79b LDSS-3C data for 8 wavelength bins for the light curve extraction using the \textit{Divide-White} normalization method.  It should be noted that the transit depth estimate for the 0.65  {$\mu$}m channels is likely somewhat low due to detector cutoff at the blue edge.  \cite{Rackham2017} also found decreased transit depths at bluer wavelengths for GJ 1214b, a sub-Neptune orbiting a M4.5 dwarf star, which they attribute to the presence of faculae on the unocculted stellar disk.  However, observations of WASP-79 indicate that its stellar activity is low.  We collected XMM-Newton observations of WASP-79 on 2017-07-18, with S/N=3.4. Its X-ray emission, $L_{\rm X}=5.7\times10^{28}$ erg/s (for a d=248 pc, c.f. GAIA DR2) yields a ratio log $L_{\rm X}/L_{\rm bol}=-5.5$, indicating a low activity level, as expected for an early F star (Sanz-Forcada et al. in prep.).  The TESS data baseline varies within 1${\sigma}$ $<$ 0.1{$\%$}, so these data do not show evidence of short-term stellar activity variations in WASP-79.  Furthermore, photometric observations of WASP-79 with the Tennessee State University C14 Automated Imaging Telescope (AIT) at Fairborn Observatory (see, e.g., \cite{Sing2015} for a description of AIT operations) show no significant brightness variability within the years 2017, 2018, and 2019.  Nor does the AIT see significant variability from year to year over the same interval to a limit of $\sim$0.005 mag, confirming the absence of longer-term activity variations.  The photometric stability of WASP-79 suggests that the decreased transit depth at shorter wavelengths is not likely to be due to inhomogeneities in the stellar photosphere.  Given the low resolution of the LDSS it is not obvious what is causing the positive slope in the spectrum at bluer wavelengths.


As discussed in Section \ref{subsec:ldss3ObsRed}, the atmospheric corrections likely do not fully account for the atmospheric dynamics during the observation, and the very deep transit depth at 0.95 {$\mu$}m is likely exaggerated by interference from H\textsubscript{2}O in Earth's atmosphere.  To account for red noise in the data, the uncertainties in the LDSS transit depth estimates are multiplied by the maximum correlated noise factor for each light curve.

Transit data for WASP-79b from the HST Space Telescope Imaging Spectrograph (STIS) instrument are currently being analyzed.  STIS provides data from 0.3 -- 1.0 $\mu$m, and these data should have smaller uncertainties than the LDSS-3C data, providing more insight into the atmospheric structure of this hot Jupiter.

\subsection{Spitzer Data} \label{sec:spitzer}

\subsubsection{Observations} \label{subsec:spitzObs}
The observations analyzed here are part of Program ID 13044 (PI: Drake Deming). The target was observed during transit with IRAC channel 1 ($3.6\, \mu{\rm m}$) and channel 2 ($4.5\, \mu{\rm m}$) \citep{FazioEtal2004apjsIRAC}.  The Astronomical Observing Requests (AOR) are  62173184 and 62173696 for channels 1 and 2 respectively.  All of these observations were carried out in sub-array mode ($32\times 32$ pixels, $39{\tt "} \times 39{\tt "}$) with a 30 minute peak-up observation preceding them.  The use of a peak up observation allows the instrument to stabilize the image on the detector `sweet spot' and decreases the likelihood of a ramp in the data \citep{Ingalls2012}.  The frame time for both observations was 2 seconds.  

\subsubsection{Methods} \label{subsec:spitzMethods}
For each AOR we began with Basic Calibrated Data (BCD) available on the Spitzer Heritage Archive.  Each BCD file contains a cube of 64 frames of $64\times 64$ pixels.  Each set of 64 images comes as a single FITS file with a time stamp corresponding to the start of the first image.  We determine the time of each frame in the set by adding the appropriate multiple of the frame time to the time stamp of the first image.  The photometric extraction was performed following the methods detailed in \cite{KilSpitzer} and \cite{KilVar} utilizing both fixed and variable apertures across a range of sizes.  Background subtraction and determination of the stellar centroid and noise pixel parameter were performed in each case.

Each transit fit was based on the model of \cite{Mandel2002}  implemented in Python by the BATMAN package \citep{Kreidberg2015}.  We assumed an orbital eccentricity of zero and used the \textit{a/R*} and \textit{cosi} values derived from the TESS data from Sectors 4 and 5.  Stellar limb darkening parameters were derived from ATLAS models and interpolated bi-linearly from tables presented in \cite{Sing2010}.  We choose to use the quadratic form and fix coefficients to [0.04735, 0.15251] and [0.0604, 0.11834] for channels 1 and 2 respectively.  
The intrapixel sensitivity variation \citep{Ingalls2012}, the change in measured flux as a function of stellar centroid position and methods of correction, are well documented \citep[e.g.][]{Ingalls2016}.  Here, we employ the Nearest Neighbors method (NNBR), otherwise known as Gaussian Kernel Regression with data \citep{Lewis2013, KilSpitzer}. 

For each AOR, the best fit values for all free parameters were initially determined using matrix inversion.  The standard deviation of the normalized residuals (SDNR) times the $\beta_{red}$ factor \citep{Gillon2010} was used as a metric for selecting the best fit out of the multiple apertures.  The results from the best fit aperture were passed to a Markov Chain Monte Carlo implemented by emcee \citep{emcee} to derive uncertainties of each free parameter.  The uncertainty on each data point in the light curve is inflated by the $\beta_{red}$ factor to account for the unresolved correlated noise.  We use a number of walkers at least twice the number of free parameters and run for $10^5$ steps per walker before testing for convergence using Gelman Rubin statistics with a threshold for acceptance of 1.01 \citep{gelman1992}.  The initial 10\% of steps for each walker are discarded to remove the `burn-in' period.  

\subsubsection{Results} \label{subsec:spitzResults}
At 4.5 $\mu$m we find a transit depth of 1.1396 {$\%$} $\pm$ 0.0103 {$\%$}.  The SDNR of this observation was 0.04875 with a $\beta_{red}$ factor of 1.09.  At 3.6  $\mu$m we find a transit depth of 1.1224 {$\%$} $\pm$ 0.0080 {$\%$} with an SDNR of 0.005505 and  $\beta_{red}$ factor of 1.41.  We find the center of transit time to occur 0.009835 $\pm$ 0.0008 days  (14.15 $\pm$ 1.15 minutes)  later than the predicted transit time  \citep{Smalley2012} in channel 1 and 0.009743 $\pm$ 0.00035 days (14.0 $\pm$ 0.5 minutes) in channel 2.  

Table \ref{tab:transitDepths} provides the wavebands, normalized transit depths, and 1$\sigma$ transit depth uncertainties for the previously-described data sets.  Table \ref{tab:transitEphs} provides the transit ephemerides and uncertainties for the TESS, HST/WFC3, and LDSS-3C observations.  We used these transit times in conjunction with the \cite{Smalley2012} ephemeris to re-compute a new ephemeris and period for WASP-79b. 


\begin{figure}
\includegraphics[width=1.0\linewidth]{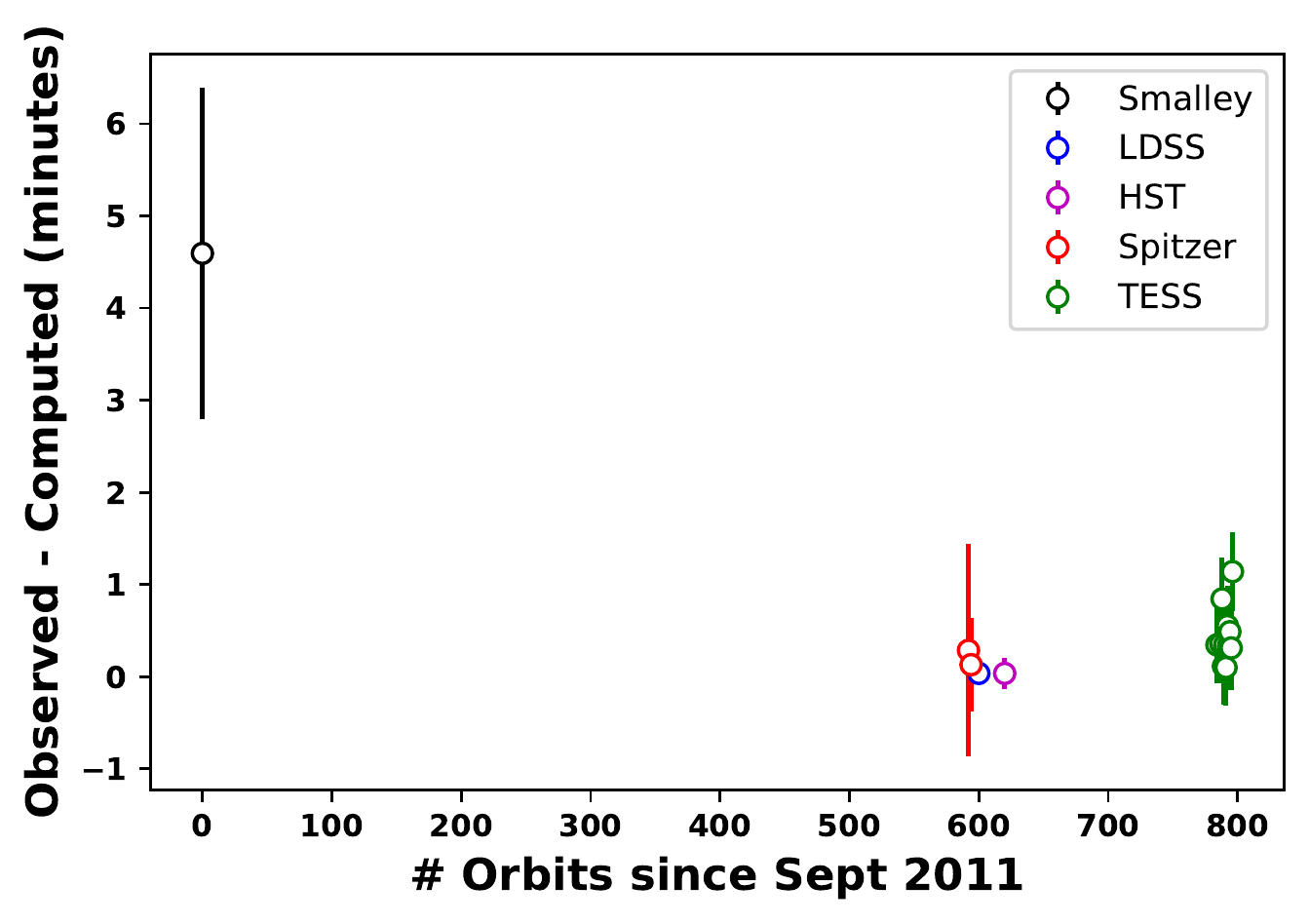}
\caption{\label{fig:transit_O-C}\small
Comparison of observed transit times with computed transit times for Smalley, WFC3, LDSS-3C, Spitzer, and TESS observations.  Computed transit times are based on the updated ephemeris and orbital period provided in Table \ref{tab:transitEphs}.}
\end{figure}

\startlongtable
\begin{deluxetable}{c|ccc}
\tablecaption{Normalized Transit Depths and Uncertainties \label{tab:transitDepths}}
\tablehead{\colhead{Instrument} & \colhead{Waveband} & \colhead{$(R_p/R_*)^2$} & \colhead{$\sigma$\textsubscript{$(R_p/R_*)^2$}} \\
\colhead{} &\colhead{($\mu$ m)} &  & \colhead{}
}
\startdata
TESS & 0.586 -- 1.031 & 1.1396 & 0.014 \\
\hline
 & 0.625 -- 0.67 & 1.0725 & 0.0316 \\
 & 0.675 -- 0.725 & 1.0955 & 0.0206 \\
 & 0.725 -- 0.757 & 1.1026 & 0.0101 \\
LDSS-3C & 0.770 -- 0.825 & 1.1175 & 0.0073 \\
 & 0.825 -- 0.875 & 1.1209 & 0.0204 \\
 & 0.875 -- 0.925 & 1.1610 & 0.0205 \\
 & 0.925 -- 0.975 & 1.2071 & 0.0215 \\
 & 0.975 -- 1.025 & 1.1282 & 0.0332 \\
 \hline
 & 1.125 -- 1.160 & 1.1486 & 0.0050 \\
 & 1.160 -- 1.195 & 1.1514 & 0.0053 \\
 & 1.195 -- 1.230 & 1.1398 & 0.0051 \\
 & 1.230 -- 1.265 & 1.1395 & 0.0047 \\
 & 1.265 -- 1.300 & 1.1385 & 0.0061 \\
 & 1.300 -- 1.335 & 1.1431 & 0.0052 \\
 & 1.335 -- 1.370 & 1.1418 & 0.0051 \\
HST/WFC3 & 1.370 -- 1.405 & 1.1634 & 0.0053 \\
 & 1.405 -- 1.440 & 1.1524 & 0.0051 \\
 & 1.440 -- 1.475 & 1.1533 & 0.0061 \\
 & 1.475 -- 1.510 & 1.1532 & 0.0053 \\
 & 1.510 -- 1.545 & 1.1412 & 0.0054 \\
 & 1.545 -- 1.580 & 1.1420 & 0.0065 \\
 & 1.580 -- 1.615 & 1.1287 & 0.0056 \\
 & 1.615 -- 1.650 & 1.1201 & 0.0072 \\
 \hline
Spitzer & 3.18 -- 3.94 & 1.1224 & 0.0080 \\
 & 3.94 -- 5.06 & 1.1396 & 0.0103 \\
\enddata
\end{deluxetable}

\startlongtable
\begin{deluxetable}{c|cc}
\tablecaption{Transit Times and Uncertainties \label{tab:transitEphs}}
\tablehead{\colhead{Instrument} & \colhead{Transit Times} & \colhead{Transit Time Error} \\
\colhead{} &\colhead{(BJD\textsubscript{TDB})} & \colhead{}
}
\startdata
Spitzer & 2457713.37538 & 8.0e-04 \\
 & 2457720.70005 & 3.5e-04 \\
 \hline
LDSS-3C & 2457742.674342 & 6.7e-05 \\
 \hline
HST/WFC3 & 2457815.92219 & 1.1e-04 \\
 \hline
& 2458412.89196 & 5.4e-04 \\
& 2458416.55480 & 2.9e-04 \\
& 2458427.54200 & 3.0e-04 \\
& 2458431.20355 & 3.1e-04 \\
& 2458434.86644 & 2.9e-04 \\
TESS & 2458438.52868 & 3.1e-04 \\
& 2458442.19138 & 2.9e-04 \\
& 2458445.85332 & 3.0e-04 \\
& 2458449.51586 & 3.3e-04 \\
& 2458453.17815 & 3.2e-04 \\
& 2458456.84066 & 3.2e-04 \\
& 2458460.50406 & 3.0e-04 \\
\hline
New Epoch & 2455545.23874 & 3.7e-04 \\
New Period (days) & 3.66239264 & 5.6e-07 \\
\enddata
\end{deluxetable}

\section{Discussion} \label{sec:discuss}

\subsection{Transmission Spectra Retrieval Analysis} \label{sec:transSpecRetAn}


We performed two atmospheric retrievals on the HST, LDSS, TESS, and Spitzer data using the ATMO code, which is described extensively in other works \citep{Amundsen2014, Tremblin2015, Tremblin2016, Tremblin2017, Drummon2016, Goyal2018, Evans2019}.  We performed a chemical equilibrium retrieval as well as a free-chemistry retrieval with FeH and H\textsuperscript{-}, as FeH is one of the few molecules likely to be found at these temperatures that has a maximum opacity at 1 {$\mu$}m \citep{Tennyson2018}.  For the stellar mass and radius, we assumed the main sequence values published by \cite{Smalley2012} -- $R_{\ast}$ = 1.64 $R_\odot$ and $M_{\ast}$ = 1.56 $M_{\odot}$ -- since their radius is consistent with that in the Gaia Data Release 2. We used a Differential-evolution MCMC to marginalize the posterior distribution \citep{Eastman2013}.  We ran twenty-two chains each for 30,000 steps and discarded the first 2\% of each chain as burn-in before combining them into a single chain.

For the model assuming chemical equilibrium, the relative elemental abundances for each model were calculated in equilibrium on the fly, with the elements fit assuming solar values and varying the metallicity ([M/H]).  However, we allowed for non-solar elemental compositions by varying the carbon, oxygen and potassium elemental abundances ([C/C\textsubscript{$\odot$}], [O/O\textsubscript{$\odot$}], [K/K\textsubscript{$\odot$}]) separately.  For the spectral synthesis, we included the spectrally active molecules of H\textsubscript{2}, He, H\textsubscript{2}O, CO\textsubscript{2}, CO, CH\textsubscript{4}, NH\textsubscript{3}, Na, K, TiO, VO, FeH, and Fe.  The temperature was assumed to be isothermal, fit with one parameter, and we also included a uniform grey cloud parameterized by an opacity and a cloud top pressure level. 

\begin{figure}[t]
\includegraphics[width=1.0\linewidth,clip]{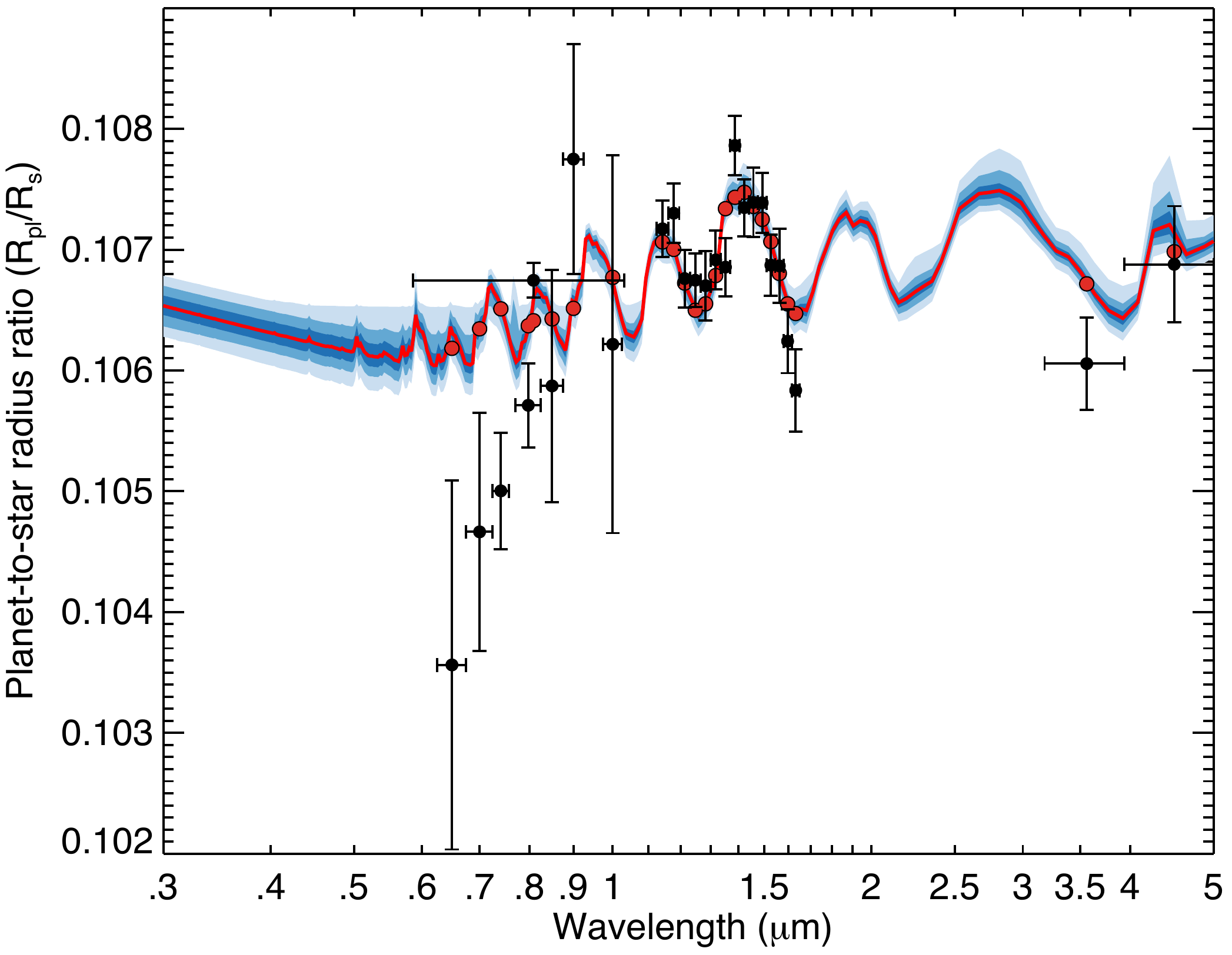}
\caption{\label{fig:atmo_retr}\small
Atmospheric spectrum from chemical equilibrium ATMO retrieval based on HST, LDSS, TESS, and Spitzer transit depth estimates.  The red line shows the best-fit model, and the blue areas indicate the 1, 2, and 3${\sigma}$ uncertainties. Due to the high cloud deck that the model is driven to by the opacity at $\sim$1{$\mu$}m, this model does not fit the decreased absorption at 1.6 {$\mu$}m or the positive slope in the bluer wavelengths.  This model had a BIC of 70.75 for 25 data points and 8 free parameters.}
\end{figure}

Figure \ref{fig:atmo_retr} shows the chemical equilibrium retrieval spectrum with the estimated transit depths.  Since the LDSS-3C spectrum for WASP-79b shows an unexpected positive slope from 0.65 {$\mu$}m to 0.8 {$\mu$}m, rather than the anticipated negative slope due to Rayleigh scattering, the model has a hard time reproducing the LDSS-3C data in the shorter wavelengths.  This retrieval is driven toward a low temperature of $\sim$800 K, which would be unexpected for this planet, as the equilibrium temperature is $\sim$1800 K. The retrieval is also driven toward high clouds by the muted 1.3 {$\mu$}m range of the HST data, which is relatively flat and high compared to the 1.4 {$\mu$}m feature, which is large and dips down comparatively far at 1.6 {$\mu$}m.  The chemical equilibrium model essentially is forced to use clouds to fit these features, though with a BIC of 70.75, this model does not provide a particularly good fit. 

For the free-chemistry retrieval, we assumed a constant abundance for each molecule that was independently fit, and we varied the H\textsubscript{2}O, CO, Na, K, VO, FeH and H\textsuperscript{-} abundances; we included only these molecules as we expect them to have strong spectral features in the wavebands corresponding to the data.  Similar to the equilibrium model, we also included a grey cloud and assumed an isothermal temperature profile.  The free-retrieval results in a better fit, with a BIC of 60.75 for the same number of data points and free parameters, as it fills in the 1.2 {$\mu$}m HST opacity, where we would expect to see a larger dip at $\sim$1 $\mu$m if water were the only absorber at these wavelengths (Figure \ref{fig:atmo_retr_piece}, \citep{Tennyson2018}).  With the opacity of FeH at $\sim$1{$\mu$}m, this model better accommodates the slope of the water feature at $\sim$1.6 {$\mu$}m as well as the diminishing opacity in the bluer wavelengths.  The H\textsuperscript{-} provides additional opacity in the 0.7 to 1.3 {$\mu$}m range, decreasing the amount of FeH in the atmosphere that is needed to reproduce the opacity in the HST data.  The H$_2$O volume mixing ratio is well-constrained to an abundance of --2.20 $\leq$ log(H\textsubscript{2}O) $\leq$ --1.55, which is 40x solar.  Similar results have been found for WASP-121b \citep{Evans2018,Evans2019}.  This model also allows for a clearer atmosphere than the chemical equilibrium model.  The temperature is still lower than that expected by equilibrium (1140 K $\pm$ 180), though the temperature uncertainties are large, and the marginalized distribution differs with the equilibrium value by less than 3-sigma confidence.

As can be seen in the posterior distribution in Figure \ref{fig:corner_plot}, water and temperature are well-constrained.  For the cloud top, we see a degeneracy between its altitude and its opacity.  We also see a degeneracy between FeH and H\textsuperscript{-}, implying an upper limit to the amount of H\textsuperscript{-} that we can expect in this atmosphere.  The upper limit on VO implies that there is no significant amount in this atmosphere. The Spitzer data weakly constrain the upper limits for CO/CO\textsubscript{2} but do not provide a lower limit.

\begin{figure}[t]
\includegraphics[width=1.0\linewidth,clip]{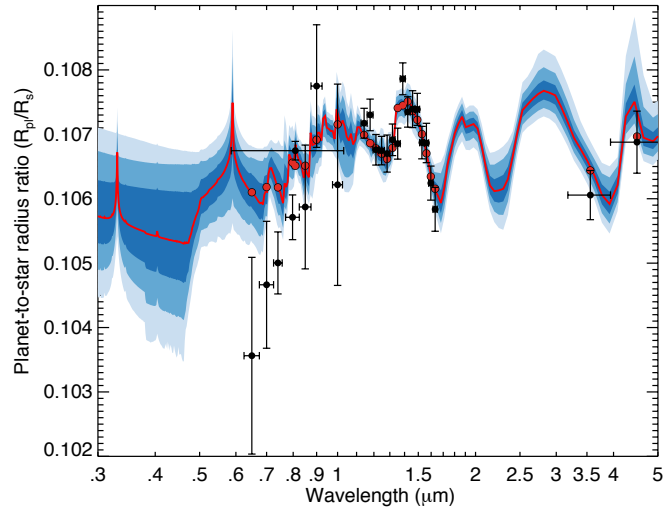}
\caption{\label{fig:atmo_retr_free}\small
Atmospheric spectrum from ATMO free-chemistry retrieval based on HST, LDSS, TESS, and Spitzer transit depth estimates.  The red line shows the best-fit model.  With the FeH and H\textsuperscript{-}, this model better accommodates the slope of the water feature at longer wavelengths as well as the diminished opacity in the bluer wavelengths.  This model also accommodates a clearer atmosphere than the chemical equilibrium model, as well as a higher temperature ($\sim$1200 K).  This model had a BIC of 60.75 for 25 data points and 8 free parameters.}
\end{figure}

While we don't spectrally resolve Na, the free-chemistry retrieval includes it because the TESS transit depth is deeper than that for LDSS-3C, and the TESS data extend into wavebands where Na features are present.  This can lift the retrieval model of the TESS data point above the LDSS-3C spectrum.  In practice, other absorbers may be causing absorption shortward of the LDSS-3C data.

\cite{Bean2018} provides the atmospheric retrieval results for WASP-79b including only the HST/WFC3 observation data with contributions from haze scattering.  Figure \ref{fig:jwst_sim} shows the retrieval spectrum with simulated JWST observation data and demonstrates the constraints that the LDSS-3C data place on the scattering slope for WASP-79b. With the large error bars of the LDSS-3C data and the precise TESS data, the LDSS-3C data do not highly constrain the retrieval, but they do help rule out large scattering slopes, as was previously thought to be likely \citep{Bean2018}

Using the methods described in \cite{Stevenson2016b}, we compute a H\textsubscript{2}O - J(H) index for WASP-79b of 0.659.  Given its temperature and log $g$, this H\textsubscript{2}O - J(H) being less than 1.0 rules out the diagonal dashed line in Figure 2 of \cite{Stevenson2016b} as a suitable boundary between clear and cloudy atmospheres and provides a better constraint on the empirical relationship between water feature strength and surface gravity.




\begin{figure}
\includegraphics[width=1.0\linewidth,trim=105 15 105 15,clip]{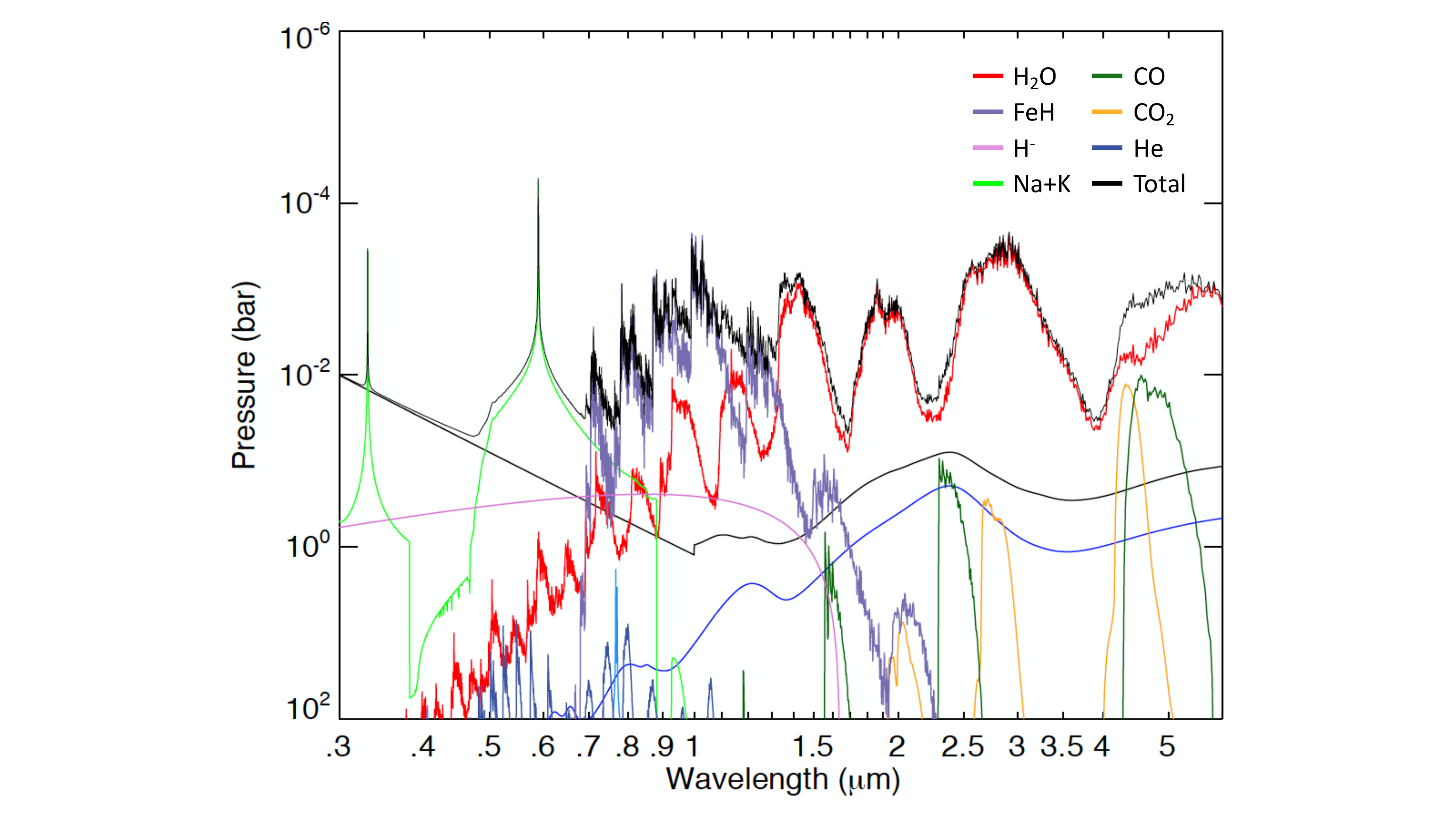}
\caption{\label{fig:atmo_retr_piece}\small
Atmospheric spectra from ATMO free chemistry retrieval showing opacity contributions from potential atmospheric components.  H\textsubscript{2}O and FeH constitute the bulk of the atmospheric opacity, with FeH providing increased opacity at $\sim$1{$\mu$}m.  The H\textsuperscript{-} provides additional opacity in the 0.7 to 1.3 {$\mu$}m range, decreasing the amount of FeH in the atmosphere that is needed.  This model allows for a clearer atmosphere than the chemical equilibrium model, as well as a higher temperature of $\sim$1200 K, which is more consistent with the expected equilibrium temperature of this planet.}
\end{figure}

\begin{figure*}[!th]
\begin{center}
\includegraphics[height=410pt,origin=c]{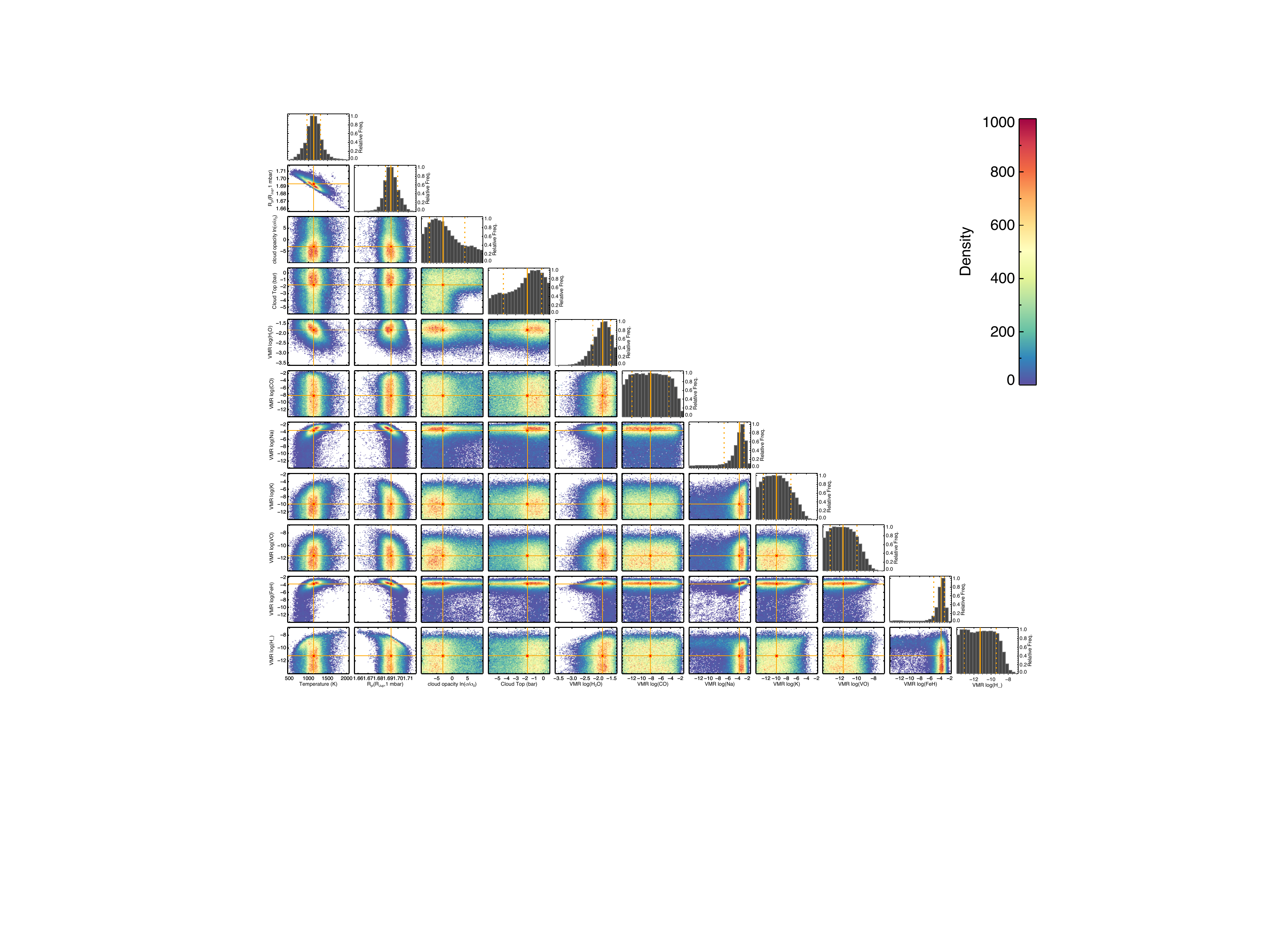}
\caption{\label{fig:corner_plot}\small
Pairs plot for the free chemistry atmospheric retrieval showing variable correlations and constraints.  The orange crosses indicate the median best fit values, and the dashed lines show the 1$\sigma$ uncertainties.  Water and temperature are well-constrained.  For the cloud top, we see a degeneracy between its altitude and its opacity.  We also see a degeneracy between FeH and H\textsuperscript{-}, implying an upper limit to the amount of H\textsuperscript{-} that we can expect in this atmosphere.  The lack of constraint on VO implies that it is not present in this atmosphere.  The combination of the deeper TESS transit depth and shallower short-wavelength LDSS3 data caused the model to include Na in the atmosphere.}
\end{center}
\end{figure*}


\subsection{JWST Expectations} \label{jwst_exp}
\begin{figure*}
\includegraphics[width=1.0\linewidth]{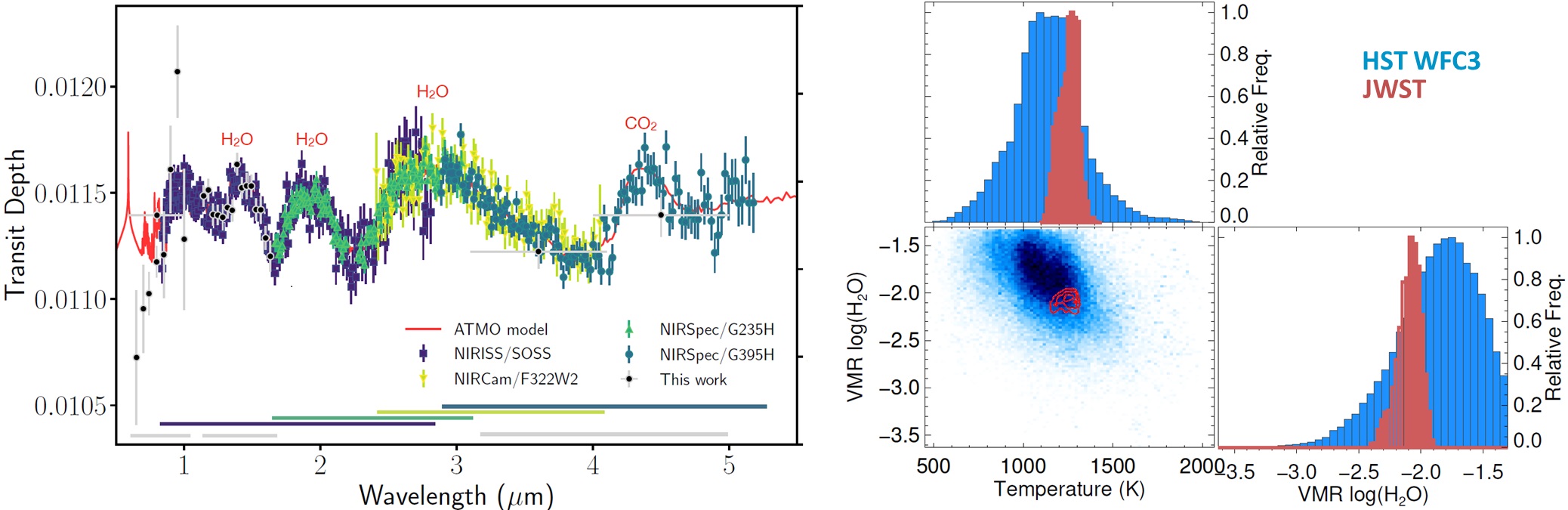}
\caption{\label{fig:jwst_sim}\small
JWST simulated observations (left) and anticipated temperature and water constraints (right) from the PanCET Program observations of WASP-79b. Left: the simulated observations were generated using Pandexo \citep{Batalha2017-PandExo}, based on the free-chemistry atmospheric retrieval spectrum and the observation data described previously. Simulated observations are shown with the estimated LDSS, TESS, HST, and Spitzer transit depths.  Results are binned to R = 100 (left).  The LDSS-3 data constrain the scattering slope, compared to Figure 7 in \cite{Bean2018}, which shows the Pandexo results for the best-fit solution for just the HST/WFC3 data with contributions from haze scattering.  Right: anticipated constraints (red) on the atmospheric temperature and water abundance compared with constraints from HST (blue).  The constraints are improved by orders-of-magnitude due to increased data resolution and the presence of multiple water features \citep{Greene2016}.}
\end{figure*}

JWST simulated observations were generated using Pandexo \citep{Batalha2017-PandExo} with the retrieval model spectrum, assuming stellar $T_{eff}$ = 6600 K, log $g$ = 4.2, and [Fe/H] = +0.03 \citep{Smalley2012}.  Figure \ref{fig:jwst_sim} shows the simulated observations for the free-chemistry retrieval model, providing an update to \cite{Bean2018}'s Figure 7 -- which was generated using just the HST data -- based on the inclusion of the LDSS, TESS, and Spitzer data in addition to the HST data.  Given these additional data, we expect to see a flatter spectrum with less pronounced Rayleigh scattering and H\textsubscript{2}O and CO\textsubscript{2} features than was originally predicted for the JWST observations.

WASP-121b \citep{Evans2016} and HAT-P-26b \citep{Wakeford2017b} also showed a similar shape in the WFC3 spectrum, with muted depth in the 1.2 -- 1.3 $\mu$m wavelength interval compared to the depth of the water feature at 1.6 $\mu$m.  Given the relatively moderate $T_{eq}$ of 990 K for HAT-P-26b, it would be unexpected for FeH to be present in its atmosphere in sufficient abundance to impact the transmission spectrum \citep{Visscher2010}, and this feature is likely best explained by a uniform scattering cloud \citep{Wakeford2017b}.  WASP-121b, however, has a $T_{eq}$ $\sim$2400 K, putting it in a temperature regime comparable to WASP-79b.  \cite{Evans2016} compared models including haze only, TiO/VO, and TiO/VO/FeH and found that the models excluding FeH could not reproduce the WFC3 transmission spectrum at wavelengths near 1.3 $\mu$m \citep{Evans2016}.  The comparable $T_{eq}$s and  similar spectrum shapes of WASP-121b and WASP-79b imply that FeH may be a spectral mechanism for both planets and should be considered in the models for similar exoplanets.



As \cite{Sing2016} note, hot Jupiters occupy a large parameter space with a wide range of gravities, metallicities, and temperatures, all of which affect a planet's atmospheric structure, circulation, and condensate formation.  It is therefore difficult to predict the spectral features of a given exoplanet. In their investigation of the influences of nonuniform cloud cover on transmission spectra, \cite{Line2016a} found that the presence of inhomogeneous clouds along the terminators of transiting exoplanets can strongly influence our interpretation of current transit transmission spectra; that a nonuniform cloud cover along the planetary terminator can influence the observed transmission spectra; and that failing to account for nonuniform cloud cover can bias molecular abundance determinations. They demonstrated that the spectrum of a globally uniform deeper cloud has a flatter shape and deeper trough than that of a nonuniform cloud cover, but that a nonuniform cloud cover spectrum was nearly identical to that produced by an atmosphere with a high mean molecular weight \citep{Line2016a}.   

However, the shape of the ingress and the egress of the transit is determined by the shape of the planetary limb and can potentially be used to constrain the cloud distribution over the planet limb and break the degeneracies between partial cloudiness and high mean molecular weight atmospheres. The shape of the residuals strongly depends on the distribution of clouds, and while the ingress and egress are symmetric in the case of polar clouds, they are antisymmetric in the case of morning clouds \citep{Line2016a}. 


These are just a few reasons why exoplanet transit transmission data are needed from JWST, a 6.5 m, space-based, near- to mid-infrared telescope. Unlike HST, which is maintained in a low Earth orbit that carries it around the globe approximately every 90 minutes, JWST will orbit at the Sun-Earth L2 point, giving it an uninterrupted view of the sky \citep{Wakeford2015}. With this uninterrupted view, JWST should be able to provide transit data with sufficiently precise timing to enable detection of clouds at the terminator. 
These more precise observations in a broader range of wavelengths will allow JWST observations of WASP-79b to contribute to the identification of clouds vs hazes in the atmosphere of this hot Jupiter.  With its muted but detectable water feature and its occupation of the clear/cloudy transition region of the temperature/gravity phase space, WASP-79b continues to represent an interesting target for the ERS program.

\section{Conclusions} \label{sec:conc}
As part of the PanCET program, we have performed a spectral analysis of the hot Jupiter WASP-79b using HST/WFC3 data (1.1 - 1.7 {$\mu$}m) and the process described in \cite{Stevenson2014a}.  We have detected a probable water feature centered at 1.4 {$\mu$}m that is consistent with the spectra of other hot Jupiters.  The LDSS-3C data (0.6 - 1.0 {$\mu$}m) are noisy, and the location of the reference star relative to the target star hindered negation of atmospheric effects occurring during the observation.  The spectrum extracted from the LDSS-3C data is therefore difficult to interpret, but overall looks relatively flat.  In conjunction with the muting of the water feature in the HST/WFC3 spectrum, this may indicate the presence of clouds in the atmosphere of this hot Jupiter, though ATMO models indicate that including the absorbers FeH and H\textsuperscript{-} provides a better fit to the data and allows for a temperature more consistent with the equilibrium temperature.  The XMM Newton, TESS, and AIT observation data indicate that the decreased transit depths in bluer wavelengths of the LDSS-3C data are not caused by stellar faculae or plage, though the low resolution of these spectral data makes it difficult to determine what may be causing these shallower transit depths.  The transit depths estimated from the TESS, LDSS, HST, and Spitzer data are all in good agreement, indicating the viability of the methods described herein.

WASP-79b represents a primary target for the PanCET program, and given the detectable water feature and the delayed launch of the JWST, it is now a primary target for the JWST Early Release Science (ERS) program \citep{Bean2018} and will be scheduled for 42 hours of JWST observation time in four different modes. These observations will provide more precise data over a broader range of wavelengths, providing a more detailed spectrum and possibly allowing for the detection of terminator clouds and/or vibrational modes of condensate species.

\acknowledgments

\textbf{Acknowledgements}  Support for program GO-14767 was provided by NASA through a grant from the Space Telescope Science Institute (STScI), which is operated by the Association of Universities for Research in Astronomy, Inc., under NASA contract NAS 5-26555.  This work is also based on observations made with the LCO Magellan Clay Telescope.  Travel to LCO/Magellan was funded by the Sagan Fellowship Program, supported by NASA and administered by the NASA Exoplanet Science Institute (NExScI).  We would like to thank Hannah Diamond-Lowe and Zach Berta-Thompson for their assistance with the LDSS-3C stretching analysis. Work done by B.M. Kilpatrick was supported by NASA Headquarters under the NASA Earth and Space Science Fellowship Program under Grant Number 80NSSC17K0484.  This portion of the work is based on observations made with the {\sl Spitzer Space Telescope}, which is operated by the Jet Propulsion Laboratory, California Institute of Technology under a contract with NASA.  Alain Lechavelier des Etangs acknowledges support from the Centre National d'{\'E}tudes Spatiale (CNES). Jorge Sanz-Forcada acknowledges funding by the Spanish MINECO grant AYA2016-79425-C3-2-P.  This project has received funding from the European Research Council (ERC) under the European Union’s Horizon 2020 research and innovation programme (project {\sc Four Aces}; grant agreement No 724427).  It has also been carried out in the frame of the National Centre for Competence in Research PlanetS supported by the Swiss National Science Foundation (SNSF).

\software{ATMO \citep{Amundsen2014, Tremblin2015, Tremblin2016, Tremblin2017, Drummon2016, Goyal2018, Evans2019},
BATMAN \citep{Kreidberg2015},
T-RECS \citep{Stevenson2016a}}
\bibliography{ms}



\end{document}